\documentclass[%
 reprint,
 superscriptaddress,
 amsmath,amssymb,
 aps,
]{revtex4-2}
\usepackage{graphicx}
\usepackage{dcolumn}
\usepackage{bm}
\usepackage{lipsum}
\usepackage{hyperref}
\usepackage{cleveref}
\usepackage{amsmath,color}
\usepackage[caption=false]{subfig}

\def\mJ{\boldsymbol{J}}
\def\mh{\boldsymbol{h}}

\begin{document}

\title{Unlearning regularization for Boltzmann Machines
}

\author{Enrico Ventura}
\affiliation{Dipartimento di Fisica, Sapienza Università di Roma, P.le A. Moro 2, 00185 Roma, Italy}
\affiliation{Laboratoire de Physique de l'Ecole Normale Sup\'erieure, ENS, Universit\'e PSL, F-75005 Paris, France}

\author{Simona Cocco}
\affiliation{Laboratoire de Physique de l'Ecole Normale Sup\'erieure, ENS, Universit\'e PSL, F-75005 Paris, France}

\author{Rémi Monasson}
\affiliation{Laboratoire de Physique de l'Ecole Normale Sup\'erieure, ENS, Universit\'e PSL, F-75005 Paris, France}

\author{Francesco Zamponi}
\affiliation{Dipartimento di Fisica, Sapienza Università di Roma, P.le A. Moro 2, 00185 Roma, Italy}
\affiliation{Laboratoire de Physique de l'Ecole Normale Sup\'erieure, ENS, Universit\'e PSL, F-75005 Paris, France}

\begin{abstract}
Boltzmann Machines (BMs) are graphical models with interconnected binary units,  employed for the unsupervised modeling of data distributions. 
When trained on real data, BMs show the tendency to behave like critical systems, displaying a high susceptibility of the model under a small rescaling of the inferred parameters. This behaviour is not convenient for the purpose of generating data, because it  slows down the sampling process, and induces the model to overfit the training-data. 
In this study, we introduce a regularization method for BMs to improve the robustness of the model under rescaling of the parameters. The new technique shares formal similarities with the unlearning algorithm, an iterative procedure used to improve memory associativity in Hopfield-like neural networks. 
We test our unlearning regularization on synthetic data generated by two simple models, the Curie-Weiss ferromagnetic model and the Sherrington-Kirkpatrick spin glass model. We show that it outperforms
$L_p$-norm schemes and discuss the role of parameter initialization. 
Eventually, the method is applied to learn the activity of real neuronal cells, confirming its efficacy at shifting the inferred model away from criticality and coming out as a powerful candidate for actual scientific implementations.   
\end{abstract}

\maketitle 


\section{Introduction}
Boltzmann Machines (BMs) \cite{hinton_unsupervised_1999, hinton_boltzmann_2007, cocco_statistical_2022} are a class of graphical models, capable of modelling data distributions through the learning of effective pairwise interactions between variables. Once properly trained {\em e.g.} through gradient ascent of the likelihood, BMs can be used to generate new data configurations, which hopefully are indistinguishable from the ones in the dataset \cite{jebara_machine_2012}. Good generative performances are, in practice, hindered by different limitations of the inferential problem.

While in theory BMs (with a sufficient number of hidden units) are universal approximators, in practical applications
one is limited by the incompatibility between the true unknown distribution of the training data and the energy-based distribution that is employed to fit them. As a consequence, the training process cannot converge to the best possible generative performance, and only selects a good approximate model among many possible similarly good choices \cite{younes_synchronous_1996}. 
Hence, regularization can be used to push the training process to yield specific desidered properties, useful for particular applications of the data analysis. 
Standard schemes for regularization include $L_2$- and $L_1$-norm penalties imposed on the pairwise interactions. However, these regularization schemes have a tendency to decrease the values of the interactions, and introduce systematic biases in the model.

Another limitation, which is intrinsic in inferential problems, is the limited availability of training data, possibly leading to the overfitting phenomenon. Informally speaking,  overfitting is an over-specialization of the model, which is too focused on the details of the specific training set, rather than capturing the broader statistical structure highlighted by the data. 
Overfitting is particularly concerning in the so-called high dimensional setting, in which the size of the model (the dimension of the data configuration) is comparable to the amount of available data. Regularization is also supposed to help in avoiding overfitting.

Due to regularization, in a statistical physics language, the inferred interactions become smaller, and the inferred model is implicitly at higher temperature. 
Correcting this bias requires introducing a temperature parameter smaller than unity when generating new data. This procedure was used in Ref.~\cite{russ2020evolution} to generate new viable protein sequences with BMs inferred from evolutionary sequence data. However, this approach is purely empirical and somewhat uncontrolled. 
It is known from the statistical physics literature that rescaling the energy function by a temperature factor, even close to one, can drastically alter the distribution properties if the system is near a phase transition point~\cite{mora_are_2011,mastromatteo_criticality_2011}.
Hence, inferring models that are robust -- i.e. generate similar data -- when the energy function is rescaled might be extremely advantageous. 

The goal of this study, motivated by the practical issues described above, is to propose a new regularization technique for BM learning that explicitly aims at increasing the robustness of the model under rescaling of its parameters, {\em i.e.} the coupling matrix $\mJ=\{J_{ij}\}$ and the local fields $\mh=\{h_i\}$. This new tool, that we named unlearning regularization, appears to be more effective than other techniques, approaching a higher robustness performance of the neural network. 
The performance of the unlearning regularization is evaluated from artificial data generated by both a Curie-Weiss (CW) and a Sherrington-Kirkpatrick (SK) models. We initially consider a small number of variables: in this way the original parameters are known, the fixed points of the learning equations are reached precisely and the useful quantities can be computed exactly. In addition, higher-dimensional and more realistic cases are also considered to validate the results. 

A particular limit of this type of regularization coincides with a thermal version of the traditional Hebbian Unlearning algorithm (HU) \cite{hopfield_unlearning_1983, van_hemmen_increasing_1990, benedetti_supervised_2022, benedetti2}. We thus specifically consider this limit and conclude that HU is able to infer the original data distribution with a very good accuracy. The performance shows an optimum in time that scales with the control parameters, as it has been observed when the algorithm is implemented in associative memory tasks \cite{nokura_paramagnetic_1996}. We conclude that HU can be interpreted as a two-steps BM learning procedure.

The structure of the article is the following. We first introduce generative modeling and Boltzmann Machines in~\cref{sec:genmod}. Some extensively employed regularization techniques (i.e. the $L_p$ methods) are then described in~\cref{sec:regs}. Section \ref{sec:u} briefly describes the HU algorithm and its traditional use in associative memory models. The new unlearning regularization is then defined in~\cref{sec:u_reg}. 
Its performance is analyzed for two different data sources in~\cref{sec:results}: a CW model (\cref{sec:CWdata}) and a SK model (\cref{sec:SKdata,sec:comp_regs0,sec:comp_regs,sec:comp_regs2}). In \cref{sec:comp_regs0,sec:comp_regs,sec:comp_regs2} we compare the new regularization method to the standard $L_p$ regularization scheme, showing that it substantially improves the robustness under rescaling of the parameters while preserving a high similarity to the original model. The analysis is performed both at convergence of the regularization algorithm and in the condition of highest similarity with the ground-truth model, obtained via early-stopping in the training, either in the small and large $N$ cases. The study of a particular limit of the regularization technique leading to HU follows in~\cref{sec:Ulimit}, which provides some insight on its behavior as a generic inference tool. Eventually, the unlearning method is tested on real biological data in \cref{sec:neurons}, confirming in practice the validity of our new regularization prescription.

\section{\label{sec:genmod} Generative Modeling}

\textit{Generative models} are a class of specific neural networks that are able to learn the probability distribution of a data-set and generate brand new data that exhibit maximum coherence with the same statistics \cite{jebara_machine_2012, cocco_statistical_2022}.\\
Consider a collection of $N$ variables in a vector $\vec{S} = (S_1,...,S_N)$. Data are $M$ realizations of such a vector grouped into a set $\{\vec{S}^{\mu}\}_{\mu = 1}^M$ and we assume that they are sampled independently from a joint distribution $P_{true}(\vec{S})$. Given $M$ data, we have access to the frequency of occurrence of a certain variable, i.e. the empirical distribution
\begin{equation}
    \label{eq:Pemp_BML}
    P_{data}(\vec{S}) = \frac{1}{M}\sum_{\mu = 1}^M\prod_{i = 1}^N \delta_{S_i^{\mu}, S_i}.
\end{equation}
Generative modeling looks for a model distribution $P_{mod}(\vec{S}|\hat{\theta})$, where the parameters $\hat{\theta}$ are inferred from the training data such that $P_{mod}$ is the closest possible to $P_{data}$. 
However, $P_{data}$ might still differ from the ground-truth distribution $P_{true}$.
As a remedy, it is useful to reduce the number of degrees of freedom of the problem by designing the model taking into account some guess we might have about $P_{true}$. For instance one might choose a \textit{graphical model}, where variables are nodes of a graph \cite{cocco_statistical_2022} with interactions to be inferred. Alternatively, one might add a prior distribution for the variables, such as a mixture of Gaussians of unknown means and unit variances~\cite{bishop_neural_1995}. 
The generative approach is largely implemented across several disciplines such as computational neuroscience \cite{cocco_neuronal_2009, schneidman_weak_2006}, bio-informatics \cite{morcos_direct-coupling_2011}, animal behaviour \cite{chen_modelling_2023}, physical simulations \cite{sims_evolving_1994}, image and text synthesis \cite{brock_large_2019, kawthekar_evaluating_2017, sohl-dickstein_deep_2015}. 
\subsection{\label{sec:BML} Boltzmann Machines}
We now describe a specific graphical model of generative neural networks, which will be particularly relevant to the present work. It is inspired by equilibrium statistical mechanics and it is called Boltzmann Machine (BM) \cite{hinton_boltzmann_2007, hinton_unsupervised_1999, jordan_graphical_2001, cocco_statistical_2022}. 

Consider a fully connected network of $N$ binary Ising variables $\vec{S} \in \{-1,+1\}^N$ with the following energy function
\begin{equation}
\label{eq:ene}
    E[\vec{S}| \mJ, \mh] = -\sum_{i, j >i}^{1,N} S_i J_{ij} S_j - \sum_{i=1}^N h_i S_i,
\end{equation}
where $J_{ij}$ are symmetric couplings and $h_i$ are the fields acting on each neuron site. Even if several types of architectures are collected under the title of BMs, we will stick to the simplest model, the one endowed exclusively with visible input neurons that interact reciprocally and are subjected to external fields. In fact, our goal will be analyzing the evolution of the interactions between real neurons, which are not necessarily present in other types of machines. \\
We know, from statistical mechanics, that such a system at equilibrium at a temperature $\beta^{-1}$ obeys the following \textit{Gibbs-Boltzmann} joint probability distribution
\begin{equation}
\label{eq:boltzpdf}
    P_{mod}(\vec{S}|\mJ, \mh, \beta) = \frac{1}{Z_{\beta}} \exp{\left(-\beta E[\vec{S}| \mJ, \mh]\right)},  
\end{equation}
with
\begin{equation}
    \label{eq:Z_boltzpdf}
    Z_{\beta} = \sum_{\vec{S}}\exp{\left(-\beta E[\vec{S}| \mJ, \mh]\right)} \ .
\end{equation}
Training a Boltzmann Machine means finding the parameters $\mJ$ and $\mh$ that minimize the distance between $P_{data}$ given by Eq.~\eqref{eq:Pemp_BML} and $P_{mod}$ given by Eq.~\eqref{eq:boltzpdf}. 
Note that as a consequence, $\beta$ can be set to one in the training without loss of generality. 
In practice, training can be achieved by minimizing the Kullback-Leibler divergence between $P_{data}$ and $P_{mod}$, i.e.
\begin{equation}
\label{eq:kull}
    \mathcal{L}(\mJ,\mh) = \sum_{\vec{S}}P_{data}(\vec{S})\log{\left( \frac{P_{data}(\vec{S})}{P_{mod}(\vec{S}|\mJ, \mh)}\right)},
\end{equation}
which is equivalent to maximizing the cross-entropy of $P_{mod}$ with respect to the empirical distribution $P_{data}$. Differentiating
\cref{eq:kull} with respect to $J_{ij}$ and $h_i$, we obtain the gradient of the loss, i.e.
\begin{equation}
    \label{eq:grad_ij}
    \nabla_{ij}\mathcal{L} = \langle S_i S_j \rangle_{mod} - \langle S_i S_j \rangle_{data},
\end{equation}
\begin{equation}
    \label{eq:grad_i}
    \nabla_{i}\mathcal{L} = \langle S_i \rangle_{mod} - \langle S_i \rangle_{data},
\end{equation}
where $\langle \hspace{0.1cm}\cdot\hspace{0.1cm}\rangle_{data}$ and $\langle \hspace{0.1cm}\cdot\hspace{0.1cm}\rangle_{mod}$ are the averages over the respective probability distributions. Therefore, the parameters can be found by iterating the following gradient descent equations
\begin{equation}
\label{eq:Jup}
    \delta J_{ij}^{(t)} = -\lambda\nabla_{ij}\mathcal{L} = \lambda \left(\langle S_i S_j \rangle_{data} - \langle S_i S_j \rangle_{mod}\right),
\end{equation}
\begin{equation}
\label{eq:hup}
    \delta h_{i}^{(t)} = -\lambda\nabla_{i}\mathcal{L} = \lambda \left(\langle S_i \rangle_{data} - \langle S_i \rangle_{mod}\right),
\end{equation}
with $\lambda$ being a small positive learning rate. 

Note that the mean and covariance over the data can be computed upstream, because they only depend on the training data-set; on the other hand, the moments of $P_{mod}$ must be re-sampled at each step of the training process, because their exact calculation would involve a sum over all possible $\vec{S}$. 
Sampling can be performed by a sufficient number of Monte Carlo chains at equilibrium at $\beta = 1$, which requires an algorithm time that is long enough to ensure ergodicity for each chain. Usually the number of chains should be of the same order of magnitude of $M$, the number of training data-points, in such a way that the statistical errors on the two averages are comparable.

Once the process converges to the fixed points of Eq.~\eqref{eq:Jup} and Eq.~\eqref{eq:hup} we obtain a \textit{moment matching} condition, i.e. the first and the second moments of the two probability distributions coincide. While, in principle, the training of a BM is a convex problem that admits a unique solution~\cite{hinton_boltzmann_2007}, in practice there are many flat directions such that some initial conditions can push the parameters closer to their optimal configuration. A common choice is, for instance
\begin{equation}
    \label{eq:J0}
    J_{ij}^{(0)} = \langle S_i S_j \rangle_{data} - \langle S_i \rangle_{data} \langle S_j \rangle_{data},     
\end{equation}
and
\begin{equation}
    \label{eq:h0}
    h_i^{(0)} = \text{atanh}\left(\langle S_i \rangle_{data}\right). 
\end{equation}

In order to measure the quality of the training of a BM one can measure whether the moment matching condition is met or not. This can be quantified by the Pearson coefficient between the moments of the $P_{mod}$ and $P_{data}$ distributions at the end of the training. Let us collect the two-points correlation matrices $\langle S_i S_j \rangle$ in a vector $\vec{c}$ where each entry runs over the indices $i, j > i$. Let us group the means $\langle S_i \rangle$ in a similar vector $\vec{\mu}$ where each entry runs over the index $i$. Then the Pearson coefficients are defined as
\begin{equation}
    \label{eq:rho_j}
    \rho_J = \frac{\sum_{i}c^{mod}_i c^{data}_i}{\sqrt{\sum_i (c^{mod}_i)^2}\sqrt{\sum_i (c^{data}_i)^2}},
\end{equation}
\begin{equation}
    \label{eq:rho_h}
    \rho_h = \frac{\sum_{i}\mu^{mod}_i \mu^{data}_i}{\sqrt{\sum_i (\mu^{mod}_i)^2}\sqrt{\sum_i (\mu^{data}_i)^2}}. 
\end{equation}

\subsection{\label{sec:regs} Regularization in Boltzmann Machines}
\begin{figure}[ht!]
\centering
\includegraphics[width=0.9\linewidth]{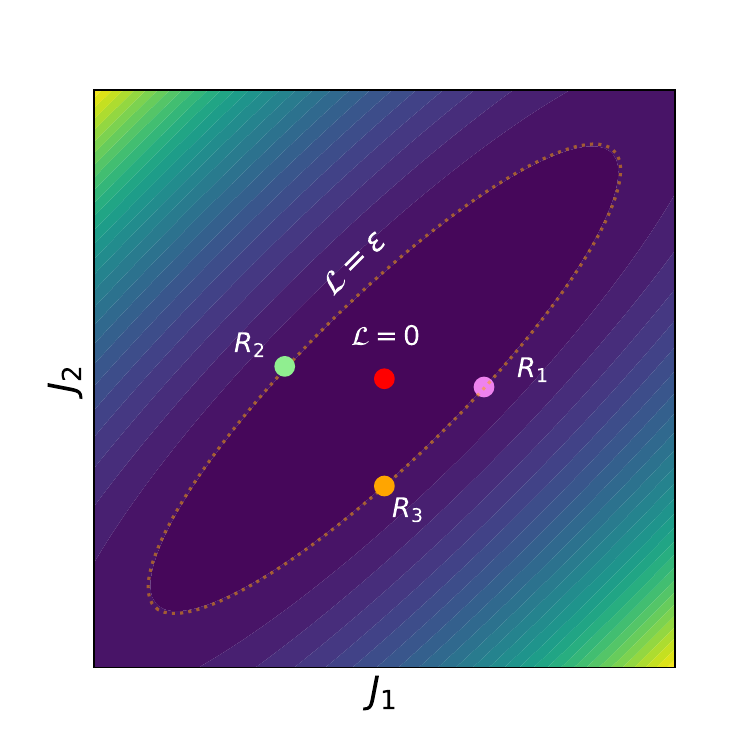}
\caption{Altimetric sketch of the typical loss function $\mathcal{L}$ minimized by BM learning, as a function of two parameters of the model. Level lines are relative to constant values of the loss. The bottom of the landscape, pointed out by the red star, is found by the unique solution (here chosen to be such that $\mathcal{L} = 0$, i.e. a perfect fit of the data). The dashed line signals all the solutions that differ by a small error $\varepsilon$ from the ground truth. Given the same distance from the bottom, colored circles indicate models found by different regularization methods.}
    \label{fig:reg_sketch}
\end{figure}
Since the loss in Eq.~(\ref{eq:kull}) is a convex function of the parameters \cite{hinton_boltzmann_2007}, there is a unique Gibbs-Boltzmann distribution that minimizes it. However, in most practical applications the data are limited and as a consequences the moments $\langle \cdot \rangle_{data}$ are only noisy estimates of the correct ones. Furthermore, the data are not generated by a Gibbs-Boltzmann distribution themselves.
Because of this, the loss landscape features many almost flat directions, corresponding to the fact that many different models can fit the data equally well, i.e. reach a loss $\mathcal{L} = \mathcal{L}_{min} + \varepsilon$ with $\varepsilon$ a small KL distance from the ground-truth. 
This idea is sketched in \cref{fig:reg_sketch}.
Regularization methods can be used to select specific models in this region of comparable loss, thus providing different inferred systems with different properties, depending on additional requirements imposed on the inferential problem. For instance, some of these models can be sparse or even contain null parameters (e.g. $L_0$, $L_1$-norm and information based regularization schemes \cite{barrat-charlaix_sparse_2021}), others are fully connected (e.g. $L_2$-norm method).  
Past literature \cite{tkacik_thermodynamics_2015, mora_are_2011, mastromatteo_criticality_2011} highlighted that critical models, i.e. inferred systems being highly susceptible to small changes in the parameters, can be attractive for BM learning. In fact, if we consider training as an homogeneous sampling of the models that minimize $\mathcal{L}$, critical models have a large basin of attraction and are thus sampled most often~\cite{mastromatteo_criticality_2011}.
Criticality can be problematic for data generation: it implies a long correlation time in the Monte Carlo dynamics, slowing down the sampling process; it makes the model susceptible under rescaling of the parameters, reducing its predictivity in real data applications. Hence, avoiding criticality might help increasing generalization.

To summarize, we define the performance of the model according to two properties: \textit{accuracy}, i.e. the capability of the model to be as similar as possible to the distribution that generated the data; \textit{robustness}, i.e. the tendency of the model not to change its typical configurations when slightly changing (e.g., rescaling by a common temperature factor) the parameters. 
The goal of this work is to introduce a new type of regularization in order to find a better compromise between these two requirements. 

We now discuss some widely used regularization methods for BMs and the way we can benchmark their accuracy and robustness.
As mentioned above, $L_p$ regularizations are certainly the most famous regularization methods, also due to their intuitive interpretation. This approach penalises high values of the parameters by adding a $L_p$ norm of the same variables in the expression of the loss function. As a consequence, the sparsity of the graph is promoted: the redundant parameters are weakened with respect to the relevant ones. The expression of the loss of a BM with $L_1$ and $L_2$ regularizations is given by the following equations
\begin{equation}
    \label{eq:L1}
    \mathcal{L}_{1} = \mathcal{L}^{BM}(\mJ, \mh) + \gamma_1\sum_{i,j>i}|J_{ij}| + \gamma_1\sum_{i}|h_i|, 
\end{equation}

\begin{equation}
    \label{eq:L2}
    \mathcal{L}_2 = \mathcal{L}^{BM}(\mJ, \mh) + \frac{\gamma_2}{2}\sum_{i,j>i}J_{ij}^2 + \frac{\gamma_2}{2}\sum_{i}h_i^2, 
\end{equation}
where $\mathcal{L}^{BM}$ is defined in Eq.~\eqref{eq:kull} and $\gamma_1,\gamma_2$ are two regularization rates. This translates into new updating rules for the parameters, i.e. 
\begin{equation}
    \label{eq:L1up}
    \delta J_{ij}^{(1)} = \delta J_{ij}^{BM} - \lambda\gamma_1\text{sign}(J_{ij}), 
\end{equation}
\begin{equation}
    \delta h_i^{(1)} = \delta h_{i}^{BM} - \lambda\gamma_1\text{sign}(h_i),
\end{equation}
\begin{equation}
    \label{eq:L2up}
    \delta J_{ij}^{(2)} = \delta J_{ij}^{BM} - \lambda\gamma_2 J_{ij},
\end{equation}
\begin{equation}
    \delta h_i^{(2)} = \delta h_{i}^{BM} - \lambda\gamma_2 h_i. 
\end{equation}

\textit{Accuracy} can be quantified by the Kullback-Leibler divergence $D_{KL}(true | mod)$ between the inferred model and the original one. Though, this quantity is not symmetric under the exchange of the two distributions, which can be inconvenient to define a distance. We can then adopt a symmetric version of the divergence as
\begin{multline}
    \label{eq:kl_sdiv}
    \small
    sD_{KL}(true,mod) =\\= D_{KL}(true|mod) + D_{KL}(mod|true).
\end{multline}
Even if numerical results do not show a significant difference between these two quantities, we will mainly adopt $sD_{KL}$ in our analysis. Sometimes, the KL divergence is not sufficient to quantify accuracy, so one must compare the model correlations and the data correlations. A detailed analysis of these will be provided in the following.  

Regarding \textit{robustness}, it is known that the susceptibility of the system to small variations of temperature can be measured through the \textit{specific heat} \cite{landau_statistical_1980}, defined as  
\begin{equation}
    \label{eq:cv}
    C_v(\beta) = -\beta\frac{\partial S_{\beta}}{\partial \beta} = \beta^2\left(\langle E^2 \rangle_{\beta} - \langle E \rangle_{\beta}^2 \right),
\end{equation}
where $S_{\beta}$ is the entropy of the model at a given inverse temperature $\beta$. 
The fact that inferred neural networks typically display a peak in $C_v$ around the value of $\beta$ used for training implies the vicinity of a critical point (the finite size of the system impedes $C_v$ to properly diverge and to show a real criticality). 
Therefore, since all parameters naturally scale with $\beta$ in Eq.~\eqref{eq:boltzpdf}, $C_v$ is a measure of the sensitivity of the model to a small perturbation of the parameters: high values of $C_v$ imply a strong variation of the model entropy under a small variation of the parameters, which suggests strong variations of the inferred statistics as well. 
Note that other kind of susceptibilities can be defined by taking derivatives of observables with respect to model parameters, but we will focus on the specific heat in the rest of our study.


\section{\label{sec:u} Unlearning regularization}

\subsection{Unlearning algorithm}

We will now describe an unsupervised algorithm employed in the realm of associative memory models that will be our main focus in the following.
Inspired by the brain functioning during REM sleep \cite{crick_function_1983}, the Hebbian Unlearning algorithm (HU) \cite{hopfield_unlearning_1983,crick_function_1983,van_hemmen_increasing_1990,taylor_unlearning_1992,benedetti_supervised_2022, wimbauer_universality_1994, Klein} is a training procedure for simple associative models. Consider a neural network having the same energy defined in Eq.~\eqref{eq:ene}. We assume the external local fields to be null, i.e. $h_i = 0$, $\forall i$. 
The associative memory task consists in retrieving a set of random neural states $\{\vec{\xi}^{\mu}\}_{\mu = 1}^{\alpha N}$, called \textit{patterns}, through a zero temperature Monte Carlo dynamics \cite{peretto_collective_1984} initialized on some corrupted versions of them. Such a rule for the dynamics reads
\begin{equation}
    \label{eq:0MCMC}
    S_i^{(t+1)} = \text{sign}\left(\sum_j J_{ij}S_j^{(t)} \right).
\end{equation}

The control parameter $\alpha$ is the \textit{load} of the model. 
Associativity is improved by achieving larger basins of attraction of the patterns. 

The training procedure starts by initializing the connectivity matrix in the Hebb's fashion as in Eq.~\eqref{eq:J0}, i.e. $J_{ij}^0 = \langle S_i S_j \rangle_{data}$, $\forall i,j$, where $\langle \cdot \rangle_{data}$ is the empirical average over the patterns. 
Then, the following routine is iterated at each step $t$:
\begin{enumerate}
    \item Initialize the network on a random neural state. 
    \item Run the zero temperature Monte Carlo dynamics, updating a randomly picked neuron at each time step according to Eq.~\eqref{eq:0MCMC}, until convergence to a stable fixed point $\vec{S}^{*}$. 
    \item Update couplings according to:
    \begin{equation}
        \label{eq:unl_rule}
        \delta J_{ij}^{(t)} = -\frac{\lambda}{N}S_i^{*} S_j^{*} \ , \hspace{1cm}J_{ii} = 0 \hspace{0.5cm}\forall i \ .
    \end{equation}
\end{enumerate}
This algorithm was first introduced in Ref.~\cite{hopfield_unlearning_1983} to prune the landscape of attractors from proliferating spurious states, i.e. fixed points of the neural dynamics not coinciding with the patterns \cite{gardner_structure_1986, amit_modeling_nodate}. This pruning action stabilizes the patterns and increases the size of their basins of attraction. The numerical analysis of Ref.~\cite{benedetti_supervised_2022} shows that HU approaches the memory performance of a maximally stable symmetric perceptron, which is considered to be a very effective model, up to a critical capacity $\alpha_c \simeq 0.6$. 
Another important aspect to underline is that HU performs at its best at a given amount of algorithmic iterations that scales with $N, \alpha$ and $\lambda$~\cite{van_hemmen_increasing_1990,benedetti_supervised_2022}. After this amount of steps the performance of the model, in terms of perfect retrieval of the patterns and size of the basins of attraction, deteriorates. This implies the necessity of applying an \textit{early-stopping} criterion for the HU algorithm to perform optimally.
We also note that HU is an \textit{unsupervised} algorithm, in the sense that it does not need to be provided explicitly with the patterns, and only exploits the information encoded in the Hebbian initialization of the couplings.

\subsection{\label{sec:u_reg} Unlearning regularization}

We now propose a new type of regularization that has the objective of imposing robustness under rescaling of the parameters, i.e. 
\begin{equation}
    \mJ \longrightarrow \beta \mJ \ , \qquad \mh \longrightarrow \beta \mh \ ,\qquad \text{with } \beta \neq 1.
\end{equation}
For the model to be robust under a redefinition of the parameters, we can add a regularization term to the traditional BM loss function that shifts the peak of the specific heat (i.e. the critical temperature) away from $\beta = 1$, where the data are generated. 
Hence, let us define the following loss function
\begin{equation}
    \label{eq:obj}
    \small
    \mathcal{L}(\mJ,\mh | a) = D_{KL}(data|1) + \left(\frac{a-1}{a}\right) D_{KL}(data|a) \ ,
\end{equation}
with
\begin{equation}
    \label{eq:crossent}
    D_{KL}(data|\beta) = \sum_{\vec{S}}P_{data}(\vec{S})\log{\left(\frac{P_{data}(\vec{S})}{P_{\beta}(\vec{S})}\right)}, 
\end{equation}
being the Kullback-Leibler divergences between the data and the model inferred at an inverse temperature $\beta$. Both the cases of positive and negative regularization factor (i.e. for $a \lessgtr 1$) will be evaluated in this work. The motivation for using the asymmetric KL distance will become clear in a few lines.   
The gradient descent equations for this loss function become
\begin{equation}
    \label{eq:Juprob}
    \begin{split}
    \delta J_{ij}^{(t)} = \lambda\big[ & a\left(\langle S_i S_j \rangle_{data} - \langle S_i S_j \rangle_{a}\right) \\ & - (\langle S_i S_j\rangle_1 - \langle S_i S_j \rangle_a )\big],
    \end{split}
\end{equation}
\begin{equation}
    \label{eq:huprob}
    \delta h_{i}^{(t)} = \lambda\left[ a\left(\langle S_i \rangle_{data} - \langle S_i \rangle_{a}\right) - (\langle S_i \rangle_1 - \langle S_i\rangle_a )\right].
\end{equation}
When $a = 1$, Eqs.~(\ref{eq:Juprob}) and (\ref{eq:huprob}) coincide with the original BM learning. In the limit $a \rightarrow 0$ one has $\langle S_i S_j \rangle_{a\to 0} \rightarrow \langle S_i \rangle_{0}\langle S_j \rangle_{0} = 0$ and Eq.~\eqref{eq:Juprob} tends to a thermal version of the HU rule performed at $\beta = 1$, i.e.
\begin{equation}
    \label{eq:Juprob1}
    \delta J_{ij}^{(t)} = -\lambda \langle S_i S_j \rangle_1,
\end{equation}
\begin{equation}
    \label{eq:huprob1}
    \delta h_{i}^{(t)} = -\lambda \langle S_i \rangle_1.
\end{equation}
Eq. (\ref{eq:Juprob1}) is similar to previous attempts in the associative memory literature \cite{nokura_paramagnetic_1996}. On the other hand, when ${a \rightarrow \infty}$, and the learning rate is redefined as $\lambda = \tilde\lambda /a \to 0$, the algorithm becomes
\begin{equation}
    \delta J_{ij}^{(t)} = \tilde\lambda\left( \langle S_i S_j \rangle_{data} - \langle S_i S_j \rangle_{\infty} \right) \ ,
\end{equation}
which also resembles the unlearning procedure: the sampling at $a=\infty$ produces fixed points of the zero-temperature dynamics, which gives back Eq.~\eqref{eq:unl_rule}.
At variance with the standard routine, there is an additional Hebbian input term $\langle S_i S_j \rangle_{data}$ which impedes the couplings to vanish at convergence.

At this point, the motivation for our choice of the loss function $\mathcal{L}(\mJ,\mh|a)$ should be clear: we search for an algorithm that interpolates between the BM learning algorithm and HU. In this way HU emerges as a particular limit of a regularization method on BMs. 

\begin{figure*}[t]
    \subfloat[]{%
        \includegraphics[width=.33\linewidth]{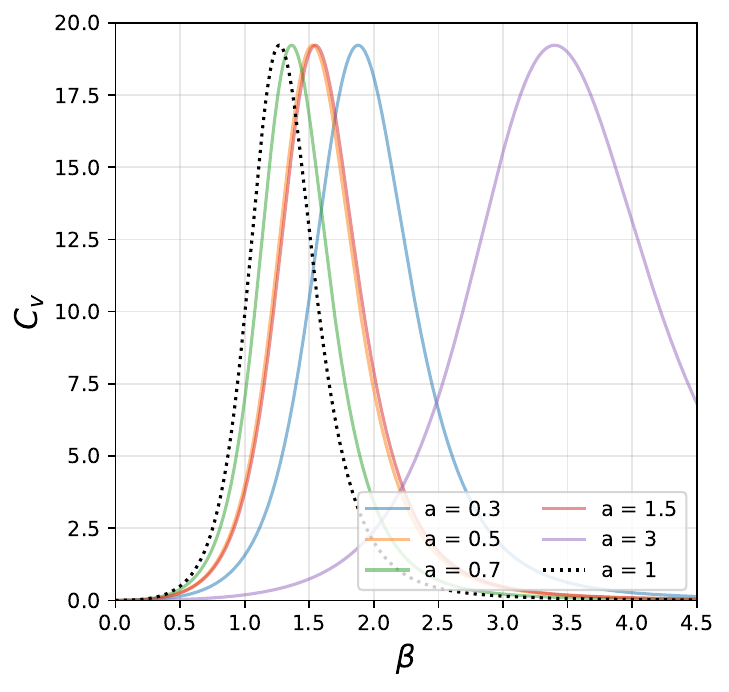}%
        \label{}%
    }\hfill
    \subfloat[]{%
        \includegraphics[width=.31\linewidth]{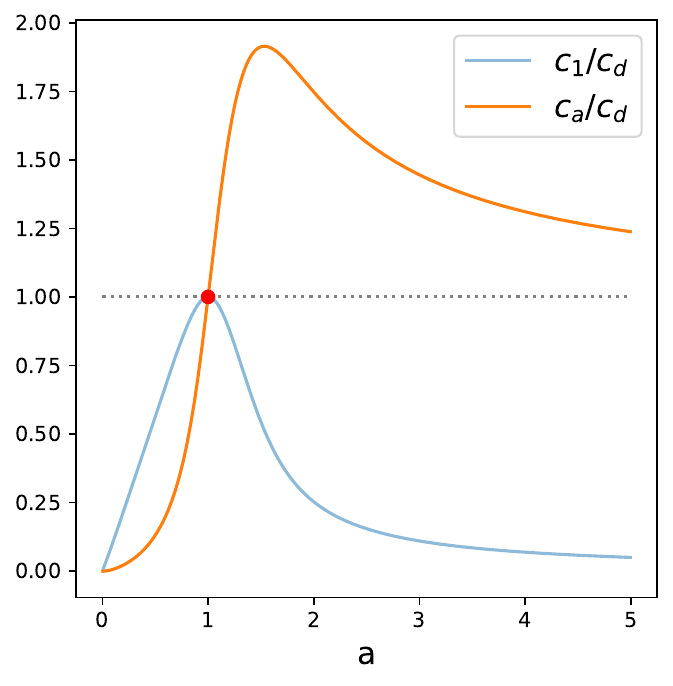}%
        \label{}%
    }\hfill
    \subfloat[]{%
        \includegraphics[width=.35\linewidth]{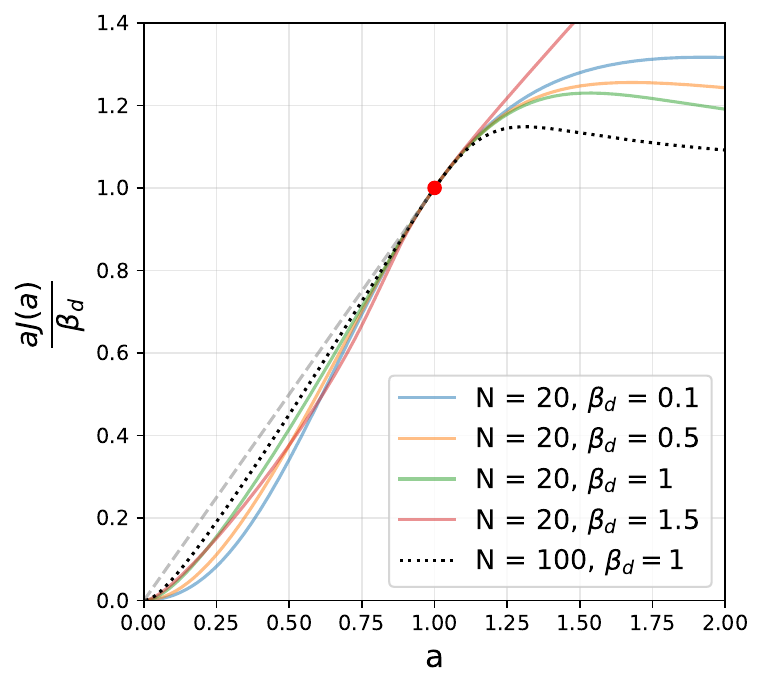}%
        \label{}%
    }
    \caption{Thermodynamic quantities computed for a small network trained on the configurations of the Curie-Weiss model with $N = 20$ spins at a given $\beta_d$. (a) The specific heat of the inferred model for $\beta_d=1$ is reported as a function of the inverse temperature $\beta$, at different values of the regularization parameter $a$. Recall that $a=1$ corresponds to standard BM learning that reproduces the original model exactly. (b) A comparison between $c_1$, $c_a$ and the empirical data correlation $c_d$ when inferred at different values of $a$. The only point where $c_1 = c_a = c_d$, i.e. $a=1$, is marked by a red circle. (c) Inferred coupling multiplied by $a/\beta_d$ as a function of $a$ at different values of $\beta_d$. The point $a=1$ where $c_1 = c_a = c_d$ is marked by a red circle.}
    \label{fig:CW}
\end{figure*}

\section{Results}
\label{sec:results}

In order to benchmark our proposed regularization scheme,
we first consider a small sized network, i.e. $N = \{18,20\}$, which allows us to consider the limit of infinite number of generated data, i.e. $M\to\infty$, and compute exactly the moments $\langle \cdot \rangle_{data}$.
We can then reach the fixed points of Eqs.~(\ref{eq:Juprob}) and (\ref{eq:huprob}) precisely, whenever such points are admitted, with no errors due to the finite sampling. We can then compute the observables $C_v$ or $sD_{KL}$ more easily, in order to determine the accuracy and robustness of the inference.
It must be underlined that our purpose in this framework is not to maximize the generalization properties of the model, because this goal would be trivially reached by the standard BM learning: it is rather to find the model that achieves the best compromise between accuracy and robustness. In the second instance, we generalize our analysis to the case of larger $N$, where quantities cannot be computed exactly. 

For simplicity of notation we rename 
\begin{equation}
    \langle S_i S_j \rangle_{data} =c_{ij}^d \ , \hspace{1.3cm}\langle S_i S_j \rangle_{\beta} = c_{ij}^{\beta} \ ,
\end{equation}
where we will mainly deal with $\beta = a$ and $\beta = 1$. The inverse temperature of the data generating model is $\beta = \beta_d$. For given $\beta_d$, we can compute $c_{ij}^d$ exactly in the limit $M\to\infty$ by exact enumeration of the model configurations.
We consider two kind of data
generating models: a mean field fully connected ferromagnetic Ising network (i.e. the Curie-Weiss model) and a fully disordered spin glass network (i.e. the Sherrington-Kirkpatrick model~\cite{sherrington_solvable_1975}), which both display a critical point in the thermodynamic limit, manifested by a singularity of the specific heat at a given critical temperature. This singularity is smoothed in finite size systems, thus appearing as a peak in specific heat. 
For both models we will choose $\beta_d$ to be smaller than the real position of the peak of $C_v(\beta)$ because sampling is more efficient in the paramagnetic phase.

\subsection{\label{sec:CWdata} Curie-Weiss model}

We first consider the problem of inferring a Curie-Weiss model (CW), that is defined by the following energy function
\begin{equation}
E_{CW}[\vec{S} | \boldsymbol{J}] = -\sum_{i,j>i}S_i J_{ij}S_j \ , \qquad J_{ij} = \frac{J}{N} \ , \forall i,j \ .
\end{equation}
This model can be fully treated analytically in the finite size case, and the procedure to compute the main quantities is reported in the following.
Since fields are zero, by construction of the model, we are interested in the evolution equation for the couplings, i.e.
\begin{equation}
    \label{eq:JuprobCW}
    \Dot{J} = \lambda\left[ a\left(c_d - c_a(t)\right) - (c_1(t) - c_a(t) )\right],
\end{equation}
where we used the fact that all elements of the matrices are identical, because each node receives the same fields from its neighbours. 
The fixed point equation for the correlation functions from Eq.~\eqref{eq:JuprobCW} is
\begin{equation}
    \label{eq:CWfixed}
    c_a^* = \frac{c_1^*}{1-a} - \frac{a}{1-a}c_d \ ,
\end{equation}
where the star indicates the value of the correlations at convergence. 
Moreover we know that 
\begin{equation}
\label{eq:CWpart}
Z_{\beta J} = \sum_m \binom{N}{\frac{N}{2}(1+m)}\exp{\left(\frac{\beta J}{2}(N m^2-1)\right)},
\end{equation}
where $m \in [-1, -1 + \frac{2}{N}, .., 1-\frac{2}{N}, +1]$
and we used
\begin{equation}\begin{split}
E_{CW}[\vec{S} | J] &= - \frac{J}{2}\left[N\left(\frac{1}{N}\sum_i S_i\right)^2 - 1\right]\\ &= - \frac{J}{2}\left(N m^2 - 1\right).
\end{split}\end{equation}
From now on, we will make use of the following notation for the correlation functions: $c_{\beta} = c(\beta J)$, $c_{d} = c(\beta_d J)$.
The second moment of the magnetization with respect to the Gibbs-Boltzmann measure $\langle m^2 \rangle_{\beta}$ can be written as
\begin{multline}
    \label{eq:msqav}
    I_N^{(2)}(\beta J) =  \frac{\sum_m m^2 \binom{N}{\frac{N}{2}(1+m)} \exp{\frac{\beta}{2} J(N m^2-1)}}{\sum_m\binom{N}{\frac{N}{2}(1+m)} \exp{\frac{\beta}{2} J(N m^2-1)}} =\\= \frac{1 + (N-1)c(\beta J)}{N}.
\end{multline}
Thus
\begin{equation}
    \label{eq:c_beta}
    c(\beta J) = \frac{N I_N^{(2)}(\beta J)-1}{N-1}.
\end{equation}
The fourth moment of the magnetization
$\langle m^4 \rangle_{\beta}$ can be written as
\begin{multline}
    \label{eq:msqsqav}
    I^{(4)}_N(\beta J) =  \frac{\sum_m m^4 \binom{N}{\frac{N}{2}(1+m)} \exp{\frac{\beta}{2} J(N m^2-1)}}{\sum_m\binom{N}{\frac{N}{2}(1+m)} \exp{\frac{\beta}{2} J(N m^2-1)}} =\\= \frac{1 + (N^3-1)k(\beta J)}{N^3},
\end{multline}
where $k(\beta J)$ is the fourth-order correlation among spins.
Eq. (\ref{eq:CWfixed}) then becomes
\begin{equation}
    \label{eq:CWfixed2}
    c(aJ) = \frac{c(J)}{1-a} - \frac{a}{1-a}c(\beta_d J),
\end{equation}
or equivalently
\begin{equation}
    \label{eq:CWfixed3}
    I_N^{(2)}(aJ) = \frac{I_N^{(2)}(J)}{1-a} - \frac{a}{1-a}I_N^{(2)}(\beta_d J),
\end{equation}
that has be to solved for $J$, given $a$ and $\beta_d$.

In the following,
we first choose $\beta_d =1$ in order to generate data from the paramagnetic phase of a network of $N = 20$ spins slightly above the critical point, since the specific heat peak of the model at $N = 20$ appears to be at $\beta > 1$.
Then, we solve Eq.~(\ref{eq:CWfixed3}) to obtain the coupling $J(a)$ of the inferred model in presence of the unlearning regularization.
Once this is done, we analyze the resulting model by computing quantities from the Gibbs-Boltzmann distribution with parameter $\beta J(a)$ rescaled by an additional temperature $\beta$.
The specific heat of the inferred model is computed as
\begin{equation}
    \label{eq:Cv_CW_analytic}
    \small
    C_v(\beta) = \left(\frac{\beta N J(a) }{2}\right)^2\left(I^{(4)}_N(\beta J(a)) - {I^{(2)}_N}^2(\beta J(a)) \right).
\end{equation}
Results are reported in \cref{fig:CW}a and show that the specific heat has a peak slightly after $\beta = 1$ that shifts progressively when $a$ is decreased as well as increased away from unity. 
In \cref{fig:CW}b we compare $c_1$ and $c_a$ with the data correlations $c_d$ as functions of $a$. There is no value of $a$ where $c_1 = c_d$ or $c_a = c_d$, signaling that the inferred model never reproduces exactly the original system, except for the trivial case of $a = 1$. Moreover, $c_1/c_d$ and $c_a/c_d$ swap when crossing $a = 1$, as one could guess by the fact that the regularization factor in \cref{eq:obj} passes from being negative to positive. 
Furthermore, in \cref{fig:CW}c we compute $J(a)$ for different values of $\beta_d$ and plot the quantity $aJ(a)/\beta_d$ as a function of $a$: when $a$ lies in between $0$ and $1$ the lines are approximately linear, signaling that $J(a)$ is nearly constant in $a$. The inferred model thus changes slowly upon varying $a$ and this property improves when $N$ is increased.

\subsection{\label{sec:SKdata} Sherrington-Kirkpatrick model: training}

Next, we consider the problem of inferring a Sherrington-Kirkpatrick model (SK), i.e.
\begin{equation}
\label{eq:HSK}
E_{SK}[\vec{S} | \boldsymbol{J}] = -\sum_{i, j>i}S_i J_{ij}S_j \ , \quad J_{ij} \sim \mathcal{N}(0,N^{-1/2}) \ .
\end{equation}
We consider a network of $N = 18$ spins with a data generation temperature $\beta_d = 0.4$. For clarity of the results, we will use only one realisation of the parameters for the SK, having a peak in the specific heat at a $\beta$ slightly larger than $\beta_d$. The parameters are initialised as $\boldsymbol{J}^{(0)} = \boldsymbol{c}_d$ and $\boldsymbol{h}^{(0)} = \boldsymbol{0} $ $\forall i$.


The fixed points of the gradient descent Eqs.~(\ref{eq:Juprob}) and (\ref{eq:huprob}) is such that the inferred model does not exactly reproduce the original system, due to the regularization. In fact, when $a \ll 1$ the stable fixed point of the equations is  $\boldsymbol{c}_1 \simeq a \boldsymbol{c}_d$ and $\boldsymbol{c}_a \simeq 0$; when $a \gg 1$, one has $\boldsymbol{c}_a = \boldsymbol{c}_d$ and $\boldsymbol{c}_1 \simeq 0$. In both scenarios the inferred model (at $a=1$) does not show a good accuracy. Note that even if for $a\gg 1$ the model scaled by $a$ fits well the data, we do not consider this as a good inferred model because it is not robust upon further lowering the temperature.

For this reason, we also consider an \textit{early-stopping} criterion for the training, such that the unlearning regularization reaches the best compromise between accuracy and robustness. 
\begin{figure*}[ht!]
    \subfloat[]{%
        \includegraphics[width=.33\linewidth]{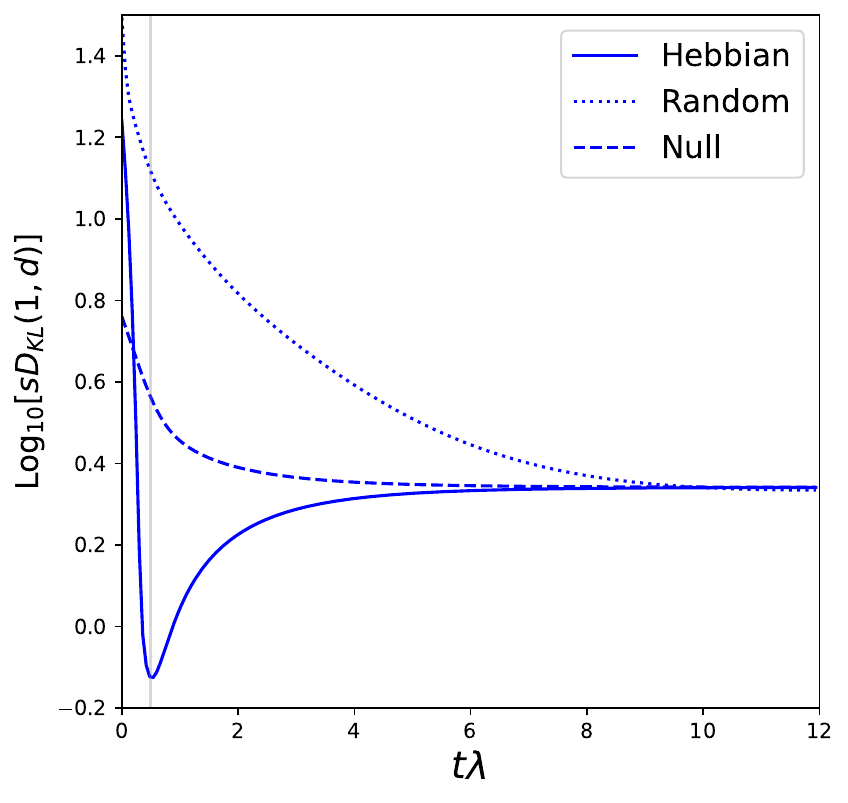}%
        \label{}%
    }\hfill
    \subfloat[]{%
        \includegraphics[width=.315\linewidth]{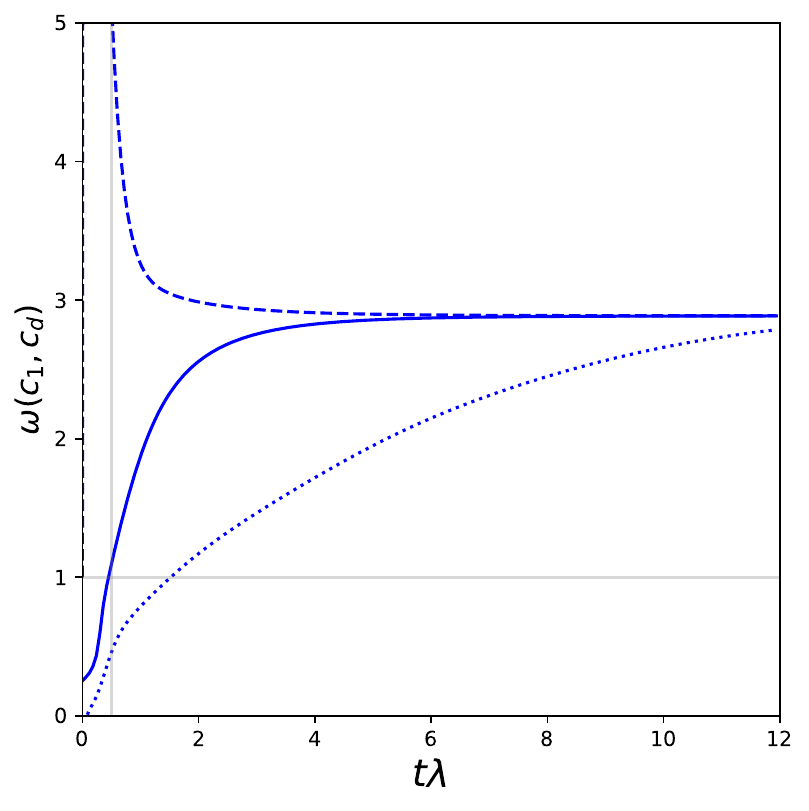}%
        \label{}%
    }\hfill
    \subfloat[]{%
        \includegraphics[width=.32\linewidth]{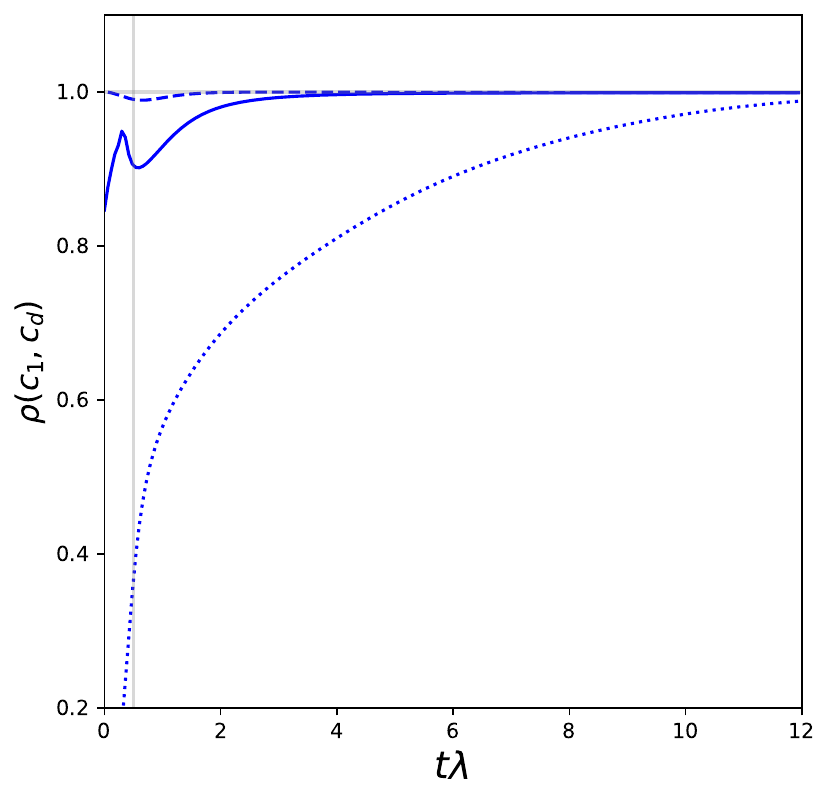}%
        \label{}%
    }
    \caption{(a) Symmetric KL divergence (in log scale), as a function of time $t\lambda$, between the inferred Gibbs-Boltzmann distribution at unitary temperature and the one describing a SK model at $\beta_d = 0.4$. (b) Slope $\omega$ obtained from the linear fit of the elements of $\boldsymbol{c}_d$ versus the elements of $\boldsymbol{c}_1$, as a function of time $t$. (c) Pearson coefficient $\rho$ between the correlation matrix $\boldsymbol{c}_{1}$ and the data correlation matrix $\boldsymbol{c}_d$ as a function of time $t$.  The \textit{full} line reports the case of Hebbian initialization of the couplings; the \textit{dotted} line corresponds to a random initialization; the   \textit{dashed} line represents the initialization to $\boldsymbol{J}^{(0)} = 0$. The gray vertical line signs the location of the minimum of $sD_{KL}$ for the Hebbian inizialization. Choice of the parameters: $N = 18$, $a = 0.3$, $\lambda = 0.06$.}
    \label{fig:SK_inits}
\end{figure*}
\begin{figure*}[ht!]
    \subfloat[]{%
        \includegraphics[width=.33\linewidth]{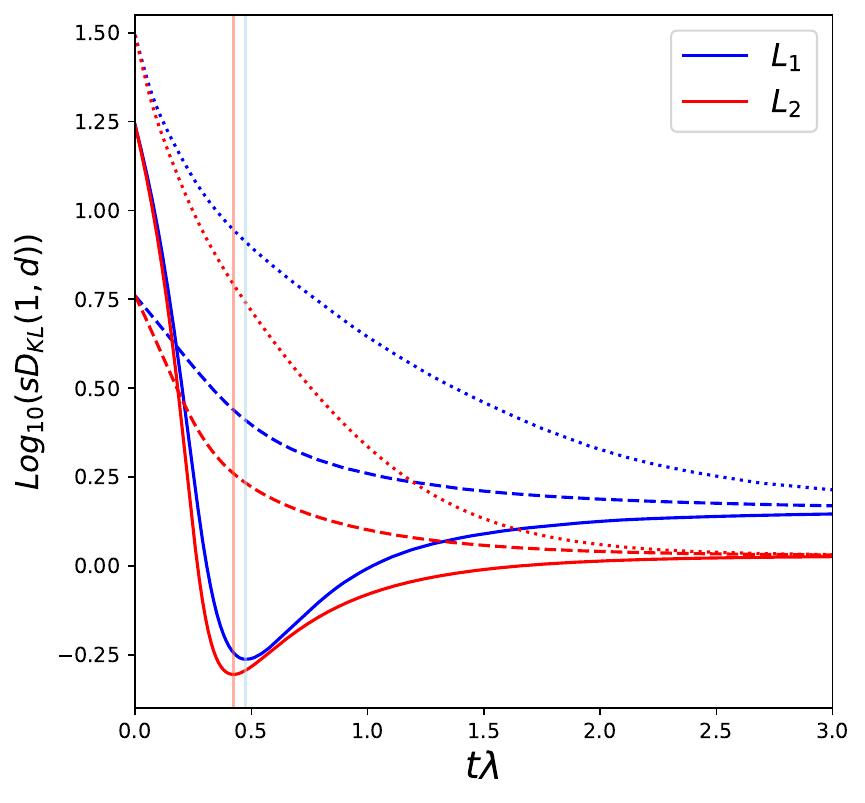}%
        \label{}%
    }\hfill
    \subfloat[]{%
        \includegraphics[width=.315\linewidth]{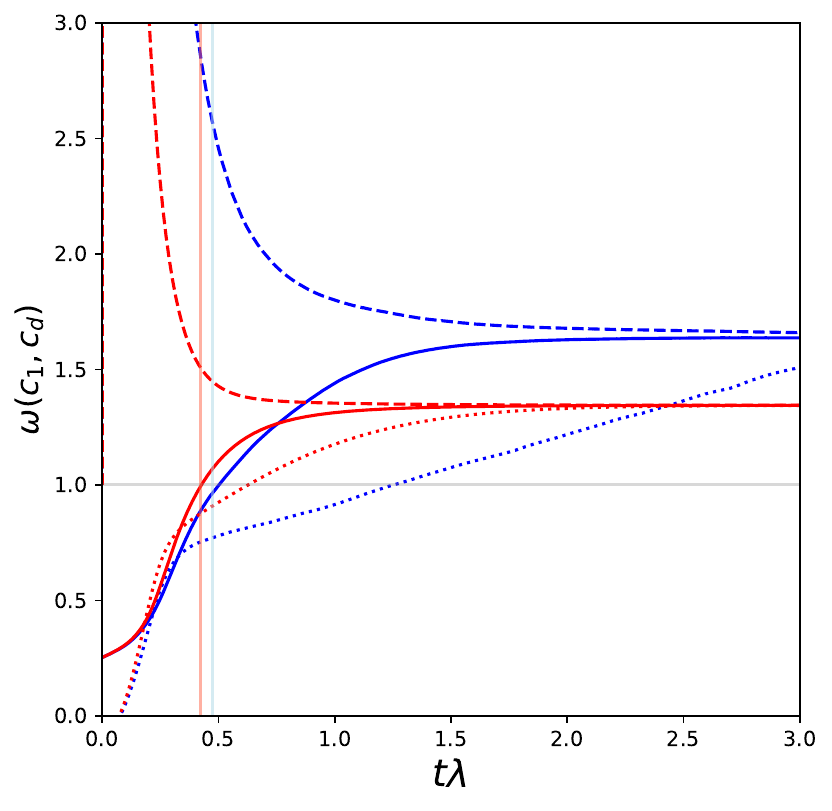}%
        \label{}%
    }
    \hfill
    \subfloat[]{%
        \includegraphics[width=.32\linewidth]{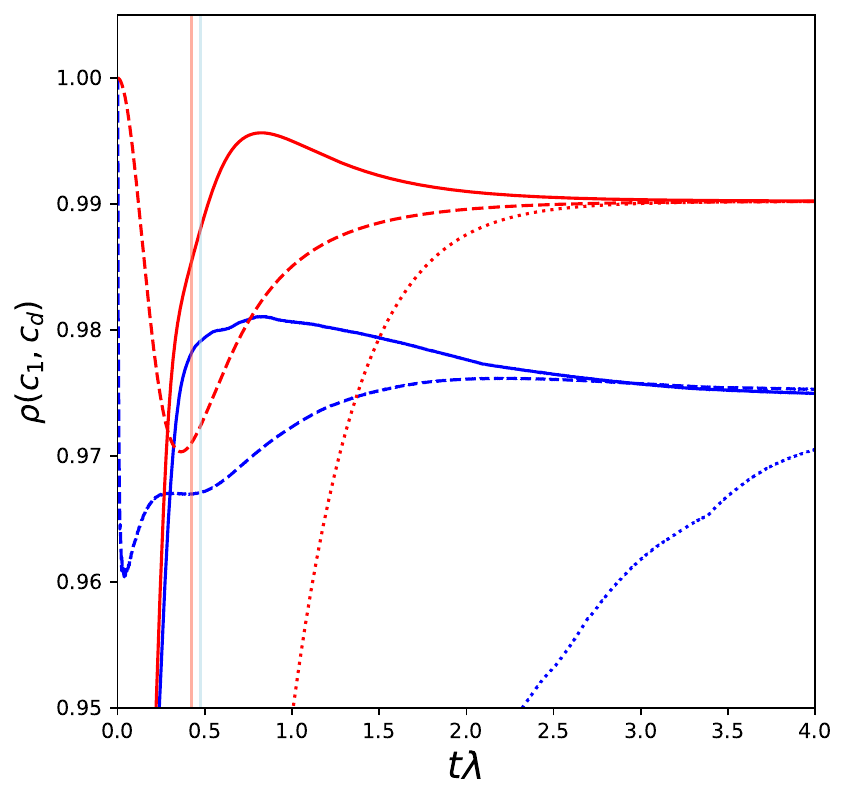}%
        \label{}%
    }
    \caption{(a) Symmetric KL divergence (in log scale), as a function of time $t\lambda$, between the Gibbs-Boltzmann distribution at unitary temperature, inferred via $L_1$ and $L_2$ regularizations, and the one describing a SK model at $\beta_d = 0.4$. (b) Slope $\omega$ obtained from the linear fit of the elements of $\boldsymbol{c}_d$ versus the elements of $\boldsymbol{c}_1$, as a function of time $t\lambda$. (c) Pearson coefficient $\rho$ between the correlation matrix $\boldsymbol{c}_{1}$ and the data correlation matrix $\boldsymbol{c}_d$ as a function of time $t\lambda$.  The \textit{full} line reports the case of Hebbian initialization of the couplings; the \textit{dotted} line corresponds to a random initialization; the   \textit{dashed} line represents the initialization to $\boldsymbol{J}^{(0)} = 0$. Light blue and red vertical lines sign the locations of the minima of $sD_{KL}$ for the Hebbian initialization, respectively in the $L_1$ and $L_2$ methods. Choice of the parameters: $N = 18$, $\beta_d = 0.4$, $\lambda_1 = \lambda_2 = 0.001$, $\gamma_1 = 0.15$, $\gamma_2 = 1$.}
    \label{fig:Lp_initcomp}
\end{figure*}
Let us compute the symmetric KL divergence between the original SK model and the inferred one at different number of training steps. Fig.~\ref{fig:SK_inits}a shows the resulting curve for a model regularized via the unlearning scheme with $a = 0.3$ and compares the standard Hebbian initialization of the couplings with two other initializations: $J_{ij}^{(0)} = 0$ and $J_{ij}^{(0)} \sim \mathcal{N}(0,N^{-1/2})$ $\forall i,j$. All the three curves reach the same final performance, by convexity of the loss function. Nevertheless, only the Hebbian initialization displays a local minimum in $sD_{KL}$ during the initial transient, which signals a high resemblance of the inferred network to the data-generating model. 

To compare the correlation matrices $\boldsymbol{c}_1$ and $\boldsymbol{c}_d$, 
a non-weighted fit of the elements of $c_d$ versus the corresponding elements of $c_1$ provides the slope $\omega(\boldsymbol{c}_1, \boldsymbol{c}_d)$, while the Pearson coefficient $\rho(\boldsymbol{c}_1, \boldsymbol{c}_d)$ measures the degree of correlation between the two matrices. A high accuracy is reached when both $\rho$ and $\omega$ are close to unity.
Note that the estimation of $\omega$ and $\rho$ carry the same computational cost, in fact at each algorithm time step we have
\begin{equation}
\small
    \label{eq:omega_comp}
    \omega^{(t)}(\boldsymbol{c}_1, \boldsymbol{c}_d) = \frac{\rho^{(t)}(\boldsymbol{c}_1, \boldsymbol{c}_d)-\mu_d\cdot\mu_1(t)}{s_{1}^2(t) - \mu_1(t)^2},
\end{equation}
where $\mu_d$ is the average over the off-diagonal elements of $\boldsymbol{c}_d$ and $\mu_1(t), s_{1}^2(t)$ are, respectively, the average of the off-diagonal elements and the average of the off-diagonal squared elements of $\boldsymbol{c}_1$ at time $t$. \\
Figures~\ref{fig:SK_inits}b and \ref{fig:SK_inits}c report the evolution of $\omega$ and $\rho$ during training. As one can notice, there is a clear correspondence between the local minimum of the symmetric KL divergence and the point where $\omega = 1$ and $\rho$ approaches a local maximum. 
It is therefore interesting to consider
 stopping the training at this point (defined in practice by $\omega=1$), in order to achieve the best accuracy. We note that this is only possible with a Hebbian initialization of the parameters; while a null inizialization never reaches $\omega = 1$, a random one reaches it but with a much lower Pearson coefficient $\rho$, resulting in poor accuracy.
Note that the correspondence between the local minimum of $sD_{KL}$ and the best match of $\boldsymbol{c}_d$ and $\boldsymbol{c}_1$ is not valid for the entire range of $a$. Specifically, for the rest of our analysis we will avoid values of $a$ that are close to unity, 
because for this value the correspondence does not hold and we cannot define a good early-stopping criterion.

We also observe that $L_p$ regularizations do not converge to the best network in terms of accuracy: even though $\rho \simeq 1$ at convergence, we find $\omega \neq 1$. Nevertheless, we found that $L_1$ and $L_2$ regularization schemes with a Hebbian initialization also display the same correspondence between a minimum in $sD_{KL}$ and the point where $\omega$ approaches unity and $\rho$ admits a local maximum in time, as shown in Fig.~\ref{fig:Lp_initcomp}.

\subsection{\label{sec:comp_regs0} SK model: results at convergence}

Before discussing the early-stopping results,
we now compare the generalization performance of the unlearning regularization with the $L_1$ and $L_2$ regularizations described in \cref{sec:regs}, at convergence. Data are generated by the same SK model with $\beta_d = 0.4$, as in \cref{sec:SKdata}.

\begin{figure*}[ht!]
\subfloat[]{%
        \includegraphics[width=.425\linewidth]{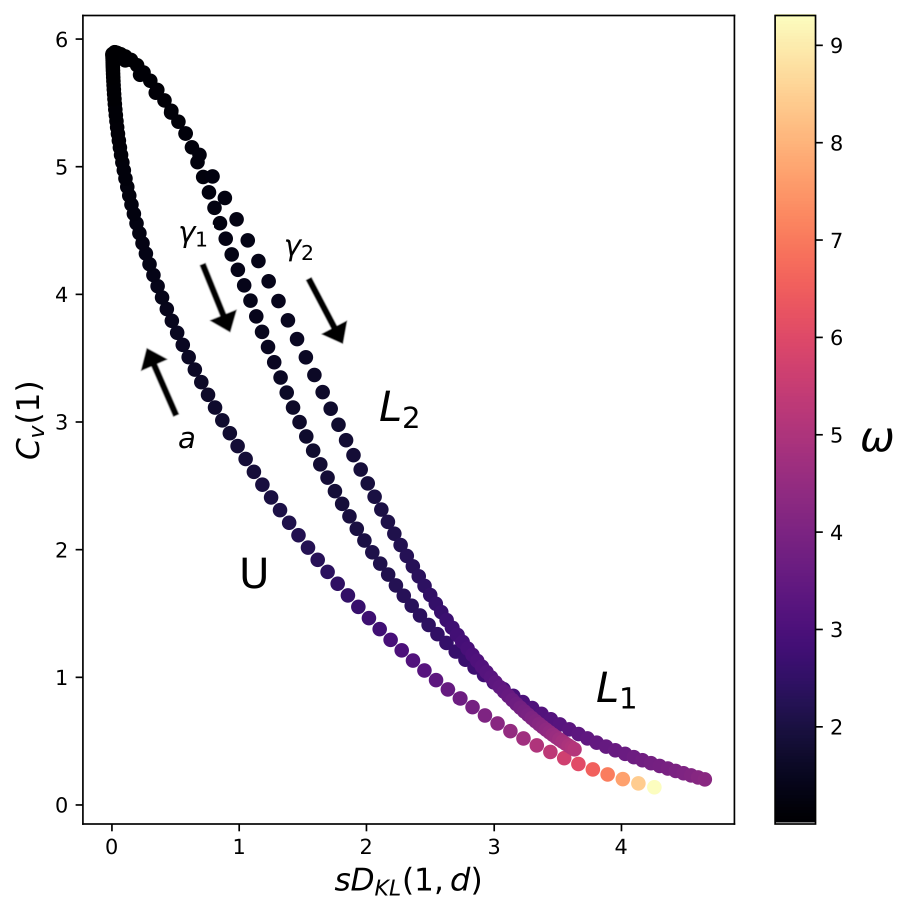}%
        \label{}%
    }\hfill
    \subfloat[]{%
        \includegraphics[width=.45\linewidth]{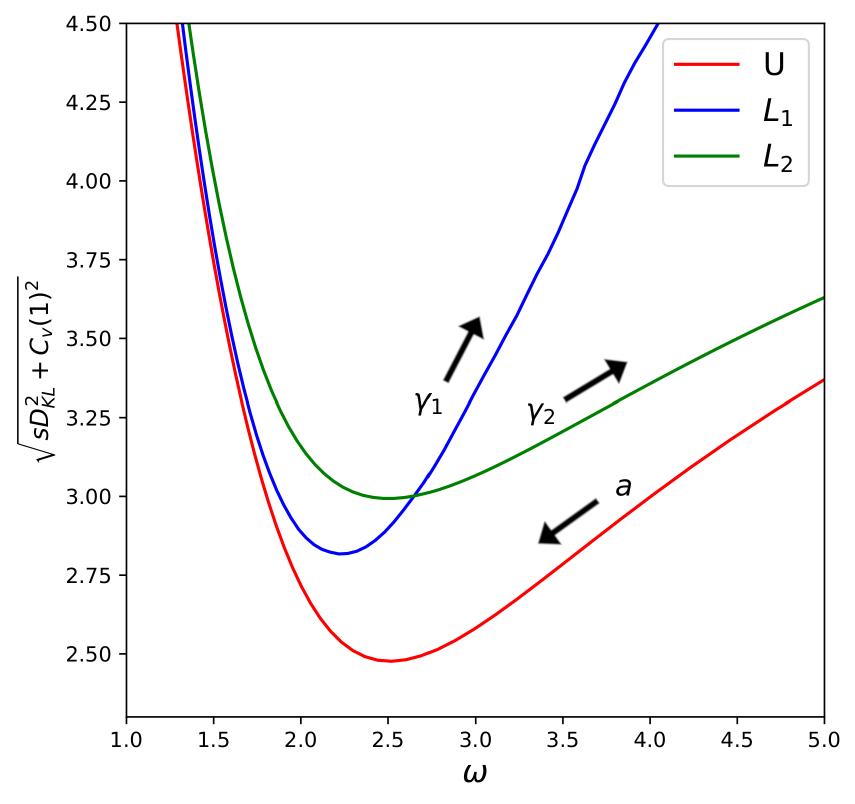}%
        \label{}%
    }
        \caption{Comparison at convergence of the performance of the unlearning, $L_1$ and $L_2$ regularization methods. (a) Symmetric Kullback-Leibler divergence $sD_{KL}$ and specific heat at unitary temperature $C_v(1)$ for different values of the slope $\omega$. (b) Distance from origin of the points in panel (a) as a function of $\omega$. Choice of the parameters: $N = 18$, $\beta_d = 0.4$, $a \in [0.1,1]$, $\gamma_1 \in [0,0.4]$, $\gamma_2 \in [0,7]$, $\lambda = 0.01$, $\lambda_1 = \lambda_2 = 0.001$.}
    \label{fig:color_comp0}
\end{figure*}

\begin{figure*}[ht!]
\subfloat[]{%
        \includegraphics[width=.425\linewidth]{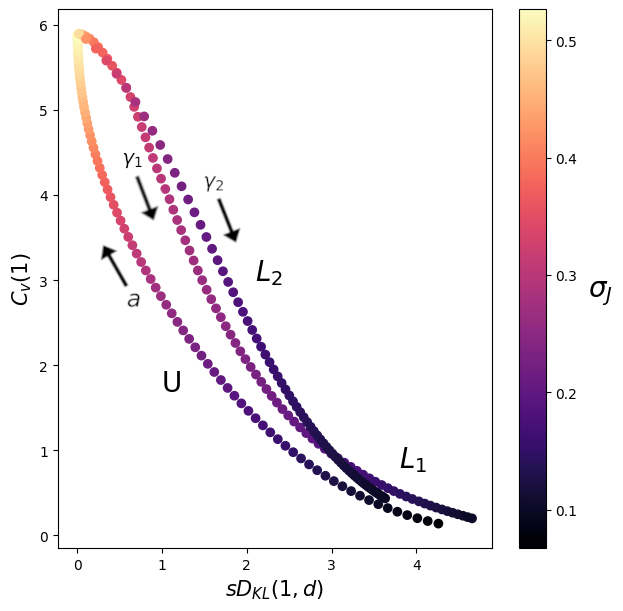}%
        \label{}%
    }\hfill
    \subfloat[]{%
        \includegraphics[width=.45\linewidth]{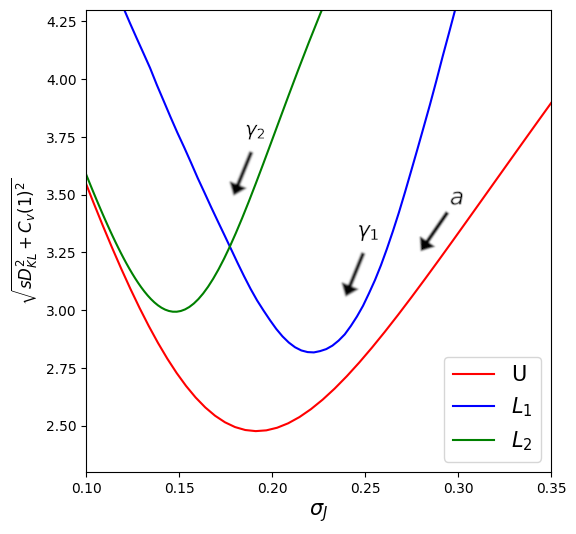}%
        \label{}%
    }
        \caption{Comparison at convergence of the performance of the unlearning, $L_1$ and $L_2$ regularization methods. (a) Symmetric Kullback-Leibler divergence $sD_{KL}$ and specific heat at unitary temperature $C_v(1)$ for different values of the standard deviations of the couplings $\sigma_J$. (b) Distance from origin of the points in panel (a) as a function of $\sigma_J$. Choice of the parameters: $N = 18$, $\beta_d = 0.4$, $a \in [0.1,1]$, $\gamma_1 \in [0,0.4]$, $\gamma_2 \in [0,7]$, $\lambda = 0.01$, $\lambda_1 = \lambda_2 = 0.001$.}
    \label{fig:color_comp1}
\end{figure*}

The couplings are optimized until convergence of the training for the three types of regularizations. Three quantities are measured after training: the symmetric Kullback-Leibler divergence $sD_{KL}(1,d)$ between the model at unitary temperature and the original SK model, the specific heat at unitary temperature $C_v(1)$ and the slope $\omega$ defined as in \cref{eq:omega_comp}. One should keep in mind that $\omega = 1$ is reached only by the trivial choice of the regularization parameters $a = 1$ and $\gamma_1 = 0$,  $\gamma_2 = 0$; any other choice of these parameters leads to a reduced accuracy, by definition of regularization.  
Measures are repeated for different choices of the regularization parameters $\{a, \gamma_1, \gamma_2\}$ and results are reported in \cref{fig:color_comp0}. 
Specifically, \cref{fig:color_comp0}a is a colour plot showing the performance of the network in the plane $(sD_{KL}, C_v(1))$, where each point is coloured according to the slope $\omega$. The three lines correspond to the three regularizations: the unlearning line is generally closer to the origin than the $L_1$ and $L_2$ ones, suggesting a better compromise between accuracy and robustness. All the lines converge to the ground-truth when $sD_{KL} = 0$, corresponding to absence of regularization ($a=1$ or $\gamma_1=0$ or $\gamma_2=0$).

To better compare the three regularization methods, \cref{fig:color_comp0}b reports the distance of the points reported in \cref{fig:color_comp0}a from the origin directly as a function of the slope $\omega$. Consistently with \cref{fig:color_comp0}a, the plot shows that the set of points relative to unlearning always stands below the $L_1$ and $L_2$ ones, while all regularizations tend to the ground truth when $\omega$ approaches unity. Different experiments with different data and temperatures $\beta_d$ show very similar results.

One can repeat the analysis by comparing the three regularization schemes at constant standard deviation of the couplings $\sigma_J$ instead of constant slope $\omega$. The results are reported in \cref{fig:color_comp1}, and the unlearning regularization once again reaches a better compromise between accuracy and robustness compared to the other regularization methods. 
Note that the $U$ and $L_2$ curves tend to coincide when $\sigma_J\to 0$, i.e. when all inferred couplings are small. This can be explained through a weak coupling expansion of the gradient equations for $U$ and $L_2$. When $\sigma_J \ll 1$ the two-point correlation functions can be Taylor expanded as
\begin{equation}
    \label{eq:weak_coup}
    c_{ij}^{\beta}(t) = \beta J_{ij}^{(t)} + \mathcal{O}({J_{ij}^{(t)}}^2).
\end{equation}
The expansion of Eq.~\eqref{eq:Juprob} leads to
\begin{equation}
    \label{eq:weak_Jup}
    \delta \boldsymbol{J}^{(t)} = \lambda\left(a \boldsymbol{c}_d -(1-a)\boldsymbol{J}^{(t)} + \mathcal{O}(a^2) \right).
\end{equation}
Since $\sigma_J$ is small when $a \ll 1$, in this limit we obtain
\begin{equation}
    \label{eq:weak_Jup2}
    \delta \boldsymbol{J}^{(t)} \simeq \lambda\left(a \boldsymbol{c}_d - \boldsymbol{J}^{(t)}\right),
\end{equation}
which at the fixed point becomes 
\begin{equation}
    \label{eq:weak_Jup_fp}
    \boldsymbol{J} = a \boldsymbol{c}_d \longrightarrow \sigma_J = a \sigma_d,
\end{equation}
where $\sigma_d$ is the standard deviation of the empirical correlation matrix. By repeating the same procedure with Eq.~\eqref{eq:L2up} one obtains
\begin{equation}
    \label{eq:weak_L2up}
    \delta \boldsymbol{J}^{(t)} = \lambda\left(\boldsymbol{c}_d +(1-\gamma_2)\boldsymbol{J}^{(t)}\right),
\end{equation}
which at the fixed point becomes 
\begin{equation}
    \label{eq:weak_L2up_fp}
    \boldsymbol{J} =  \frac{\boldsymbol{c}_d}{\gamma_2 - 1} \longrightarrow \sigma_J = \frac{\sigma_d}{\gamma_2 -1},
\end{equation}
for $\gamma_2 > 1$, which is justified by the fact that $\sigma_J$ is small when $\gamma_2$ is sufficiently large. Numerics support the fact that, when $\sigma_J \ll 1$ and $\sigma_J$ is the same in \cref{fig:color_comp0} for both $U$ and $L_2$, then $a(\gamma_2 -1) = 1$. Our results show that the HU algorithm, which derives from the unlearning regularization when $a \rightarrow 0$, can also be obtained via a $L_2$ regularization with a large $\gamma_2$.

  
\subsection{\label{sec:comp_regs} SK model with early-stopping}

\begin{figure}[t]
\centering
\includegraphics[width=1\linewidth]{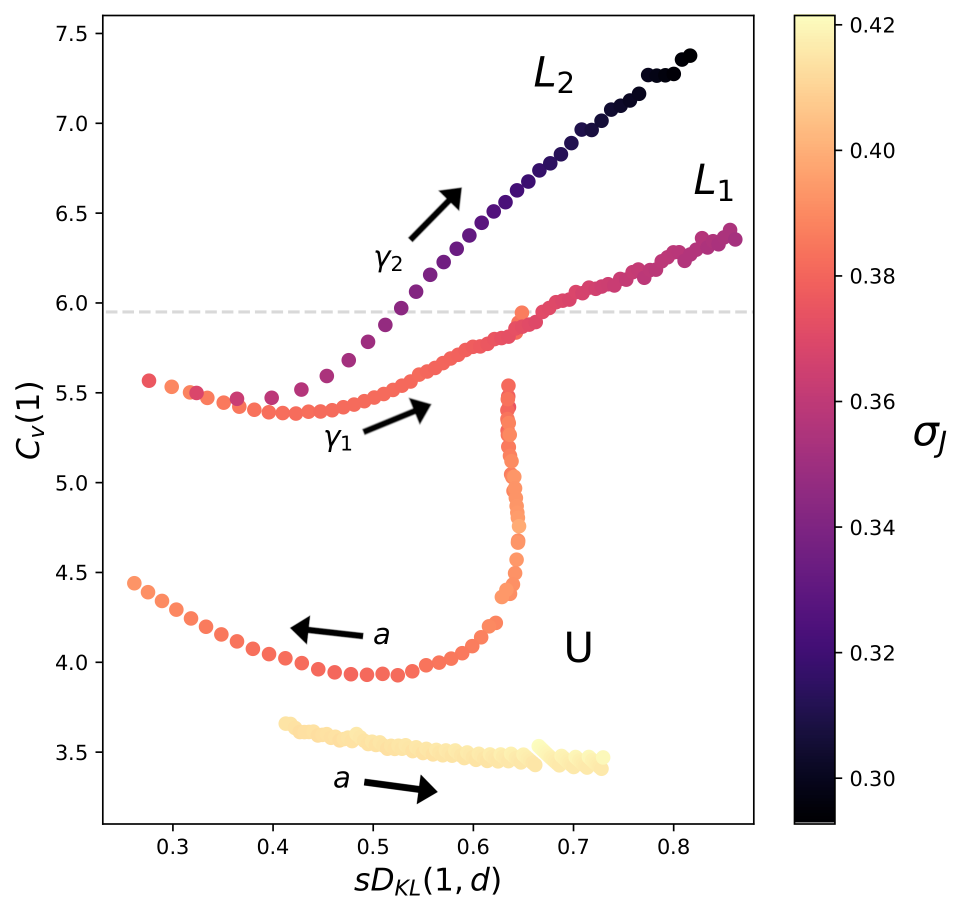}
\caption{Comparison at early-stopping of the performance of the unlearning, $L_1$ and $L_2$ regularization schemes. Each point has two coordinates: the symmetric Kullback-Leibler divergence $sD_{KL}$ and specific heat at unitary temperature $C_v(1)$. The colorbar relates each point to the standard deviation of the couplings $\sigma_J$. Arrows display the increase in the regularization rates of the algorithms. The dashed line indicates the specific heat of the original SK model that generated the data. Choice of the parameters: $N = 18$, $\beta_d = 0.4$, $a \in [0, 0.63]$ in the U upper branch and $a\in[1.9,3]$ in the U lower branch, $\gamma_1 \in [0.5,0.4]$, $\gamma_2 \in [0.3,4]$, $\lambda = 0.01$, $\lambda_1 = \lambda_2 = 0.001$.}
    \label{fig:color_comp}
\end{figure}

The early-stopping criterion described in \cref{sec:SKdata} is now employed to improve the performance of the inferred model. The training is stopped when $\omega \simeq 1$. For the unlearning regularization, we consider a range of values for $a$ such that there is a good correspondence among the minimum of $sD_{KL}$ and $\omega, \rho$ reaching unity: outside the considered interval these three quantities might not behave consistently, and we could not define a good early-stopping criterion.
Measures of the symmetric Kullback-Leibler divergence $sD_{KL}(1,d)$, the specific heat at unitary temperature $C_v(1)$ and the standard deviation of the couplings $\sigma_J$ are repeated for different choices of the regularization parameters $\{a, \gamma_1, \gamma_2\}$ and results are reported in \cref{fig:color_comp} by a colour plot showing the performance of the network in the plane $(sD_{KL}, C_v(1))$, where each point is coloured according to the standard deviation $\sigma_J$.
The specific heat of the original SK model is reported as a dashed line in the plot. The figure shows three lines relative to the three regularizations: at equal value of $sD_{KL}$, the unlearning regularization has much lower $C_v(1)$ than the $L_1$ and $L_2$ regularizations, signaling a higher robustness at comparable accuracy. We observe a range of values of $\gamma_1$ and $\gamma_2$ for which the $L_1$ and $L_2$ regularizations reduce the specific heat in $\beta = 1$, while the remaining choices for the regularizers do not allow to obtain a higher robustness with respect to the original SK model, i.e. $C_v(1)$ is increased. Different calculations with different data and temperatures $\beta_d$ produce very similar results. 

\begin{figure}[t]
\centering
\includegraphics[width=1\linewidth]{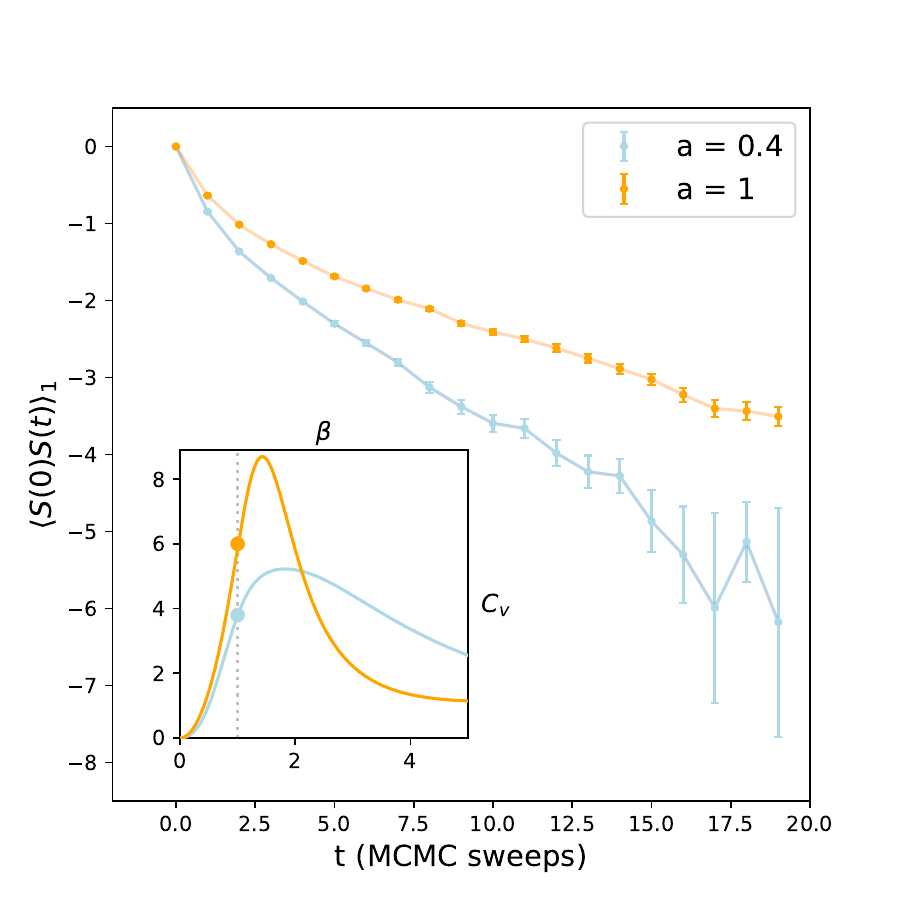}
\caption{Time correlation between network configurations sampled by a MCMC at unitary temperature when $a = 0.4$ and $a = 1$ (i.e. the original SK model), in semi-log scale. Errorbars are the propagated standard deviations of the measures. The parameters are found by the early-stopping procedure described in the text. The sub-panel reports the specific heat obtained for the same choices of $a$. Choice of the parameters: $N = 18$, $\beta_d = 0.4$, $\lambda = 0.01$.}
    \label{fig:Ct}
\end{figure}

Finally, we show that the unlearning regularization helps generating new data. Since the inferred model is less critical than the original one, the correlation between network configurations sampled at different time steps of a Monte Carlo chain at unitary temperature are smaller. Fig.~\ref{fig:Ct} shows a comparison between the time-correlation functions corresponding to the inferred system at $a = 0.4$ and the original SK model. The inset displays the behaviour of the specific heat $C_v$ for the same two models.   

\subsection{\label{sec:comp_regs2}SK model with $N \gg 1$}

In this section, we test the method on an instance with
larger number $N$ of variables, showing a good efficiency of the Unlearning method, both run until convergence and early-stopped.

A number $M = 10^4$ of data-points are sampled from a SK model with $N = 200$ spins in equilibrium at a temperature $\beta_d = 0.4$ (i.e. in the paramagnetic phase). The network is then initialized in Hebbian fashion, i.e. $\boldsymbol{J}^{(0)} = \boldsymbol{c}_d$ and $\boldsymbol{h}^{(0)} = \boldsymbol{0}$. The model is trained via Unlearning, $L_1$ and $L_2$ regularization methods and the specific heat $C_v(1)$, the slope $\omega(\boldsymbol{c}_1,\boldsymbol{c}_d)$ and the Pearson coefficient $\rho(\boldsymbol{c}_1,\boldsymbol{c}_d)$ are measured through MCMC sampling at unitary temeperature. Two different training prescriptions have been adopted for the study: the algorithm is executed until convergence of equations (\ref{eq:Juprob}), (\ref{eq:huprob}); training is early-stopped as soon as $\omega \simeq 1$. \\

\begin{figure*}[ht!]
\subfloat[]{%
        \includegraphics[width=.49\linewidth]{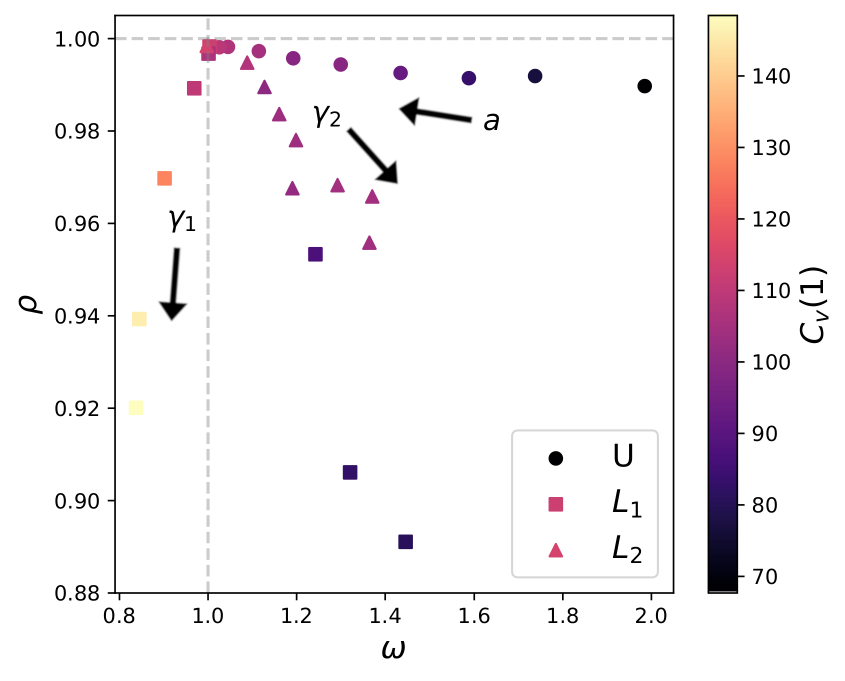}%
        \label{}%
    }\hfill
    \subfloat[]{%
        \includegraphics[width=.4\linewidth]{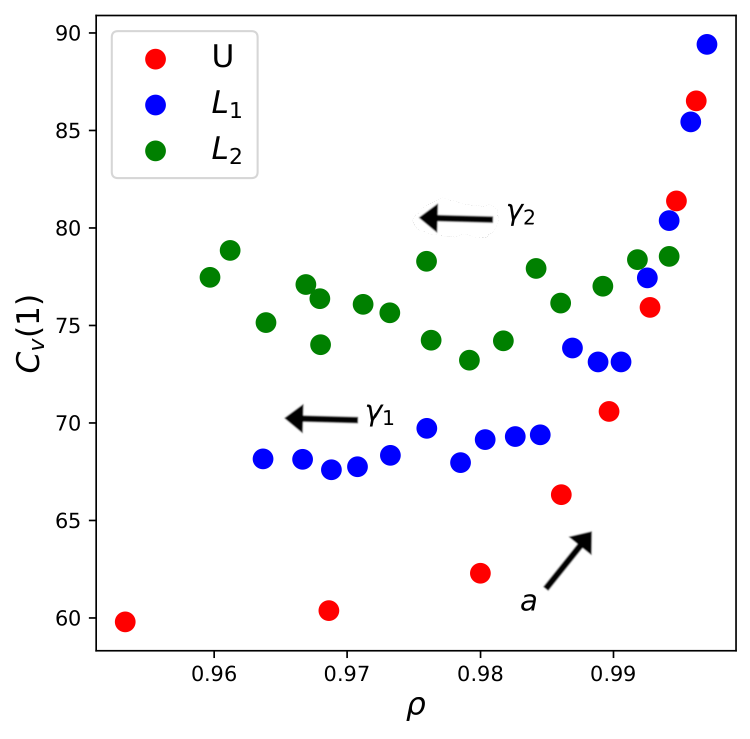}%
        \label{}%
    }
        \caption{Comparison of the performance of the unlearning with $L_1$ and $L_2$ regularization methods after training with $M = 10^4$ configurations of a SK model with $N = 200$ spins and inverse temperature $\beta_d = 0.4$. (a) The Pearson coefficient $\rho(c_1,c_d)$ is plotted as a function of $\omega(c_1,c_d)$ for different values of specific heat at unitary temperature $C_v(1)$. The dashed lines represent the best (i.e. unitary) values for $\rho$ and $\omega$, that meet in the non-regularized performance. Choice of the parameters: $\lambda = \lambda_1 = \lambda_2 = 10^{-2}$ $a \in [0.45, 1]$,  $\gamma_1 \in [0,0.09]$,  $\gamma_2 \in [0,4.5]$. (b) $C_v(1)$ as a function of $\rho(c_1,c_d)$ when the algorithm is early stopped at $\omega \simeq 1$. Measures are derived from one training instance, since they resulted stable enough over different repetitions. Choice of the parameters: $\lambda = \lambda_1 = \lambda_2 = 10^{-3}$ $a \in [0.45, 0.9]$,  $\gamma_1 \in [0.01,0.18]$,  $\gamma_2 \in [0.5,9]$.}
    \label{fig:SK_largeN}
\end{figure*}

Figure \ref{fig:SK_largeN} depicts the values of $\rho$ as a function of $\omega$ and $C_v(1)$ as achieved through the three regularization methods for various values of the control parameters $a, \gamma_1, \gamma_2$. At convergence (see \cref{fig:SK_largeN}a), the Unlearning algorithm appears to approach a better accuracy performance (i.e. in terms of both $\rho$ and $\omega$) together with a lower value of the specific heat.
Even after early-stopping (see \cref{fig:SK_largeN}b) the same behaviour emerges: there is a range in $a$ where, given the same degree of accuracy, the model achieves a smaller specific heat with respect of the other regularization methods. We also stress that, by contrast with the rest of our analysis, this experiment provides for a specific heat $C_v(1)$ obtained at early-stopping which is lower than the one reached at convergence. This aspect might be due to the Hebbian initialization of the parameters in a regime which is very far from the perfect-sampling one that we considered before. 

It should be noted that the standard deviation $\sigma_J$ of the couplings is chosen to have the same order of magnitude, across all three regularization methods. 
Both in the converged and early-stopped scenario, the Unlearning algorithm is superior to the $L_p$ regularizations, confirming the results obtained at small $N$ (i.e. the perfect sampling framework).
Moreover, the early-stopping technique used in this work, which relies on the Hebbian initialization of the network, results handily applicable for larger dimensions due to its reduced computational cost, hence providing an effective training recipe in terms of both accuracy and robustness. 

\section{\label{sec:Ulimit} The Hebbian Unlearning limit}

\begin{figure*}[t]
    \subfloat[]{%
        \includegraphics[width=.33\linewidth]{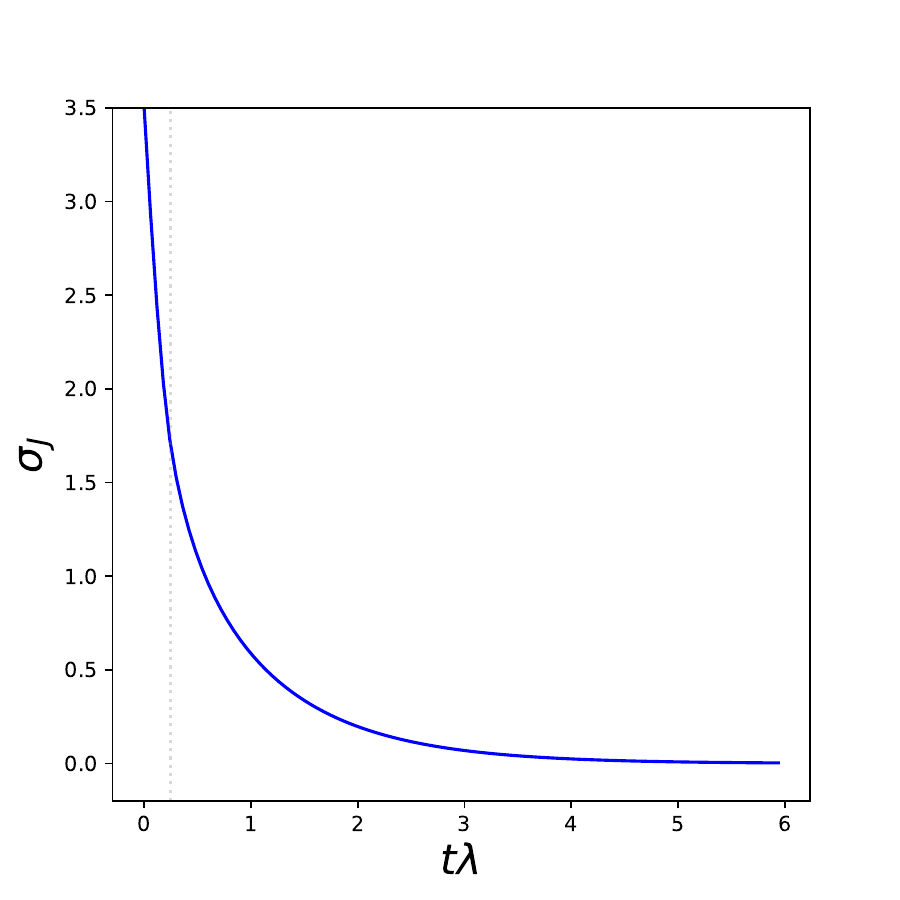}%
        \label{}%
    }\hfill
    \subfloat[]{%
        \includegraphics[width=.33\linewidth]{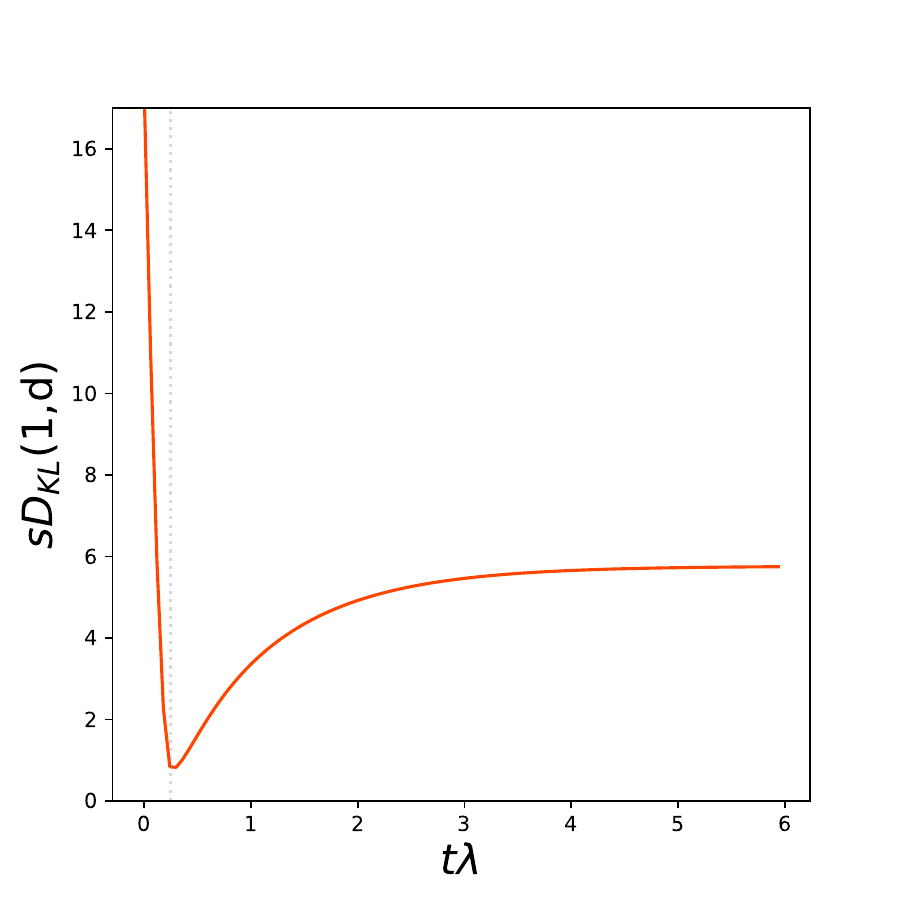}%
        \label{}%
    }\hfill
    \subfloat[]{%
        \includegraphics[width=.33\linewidth]{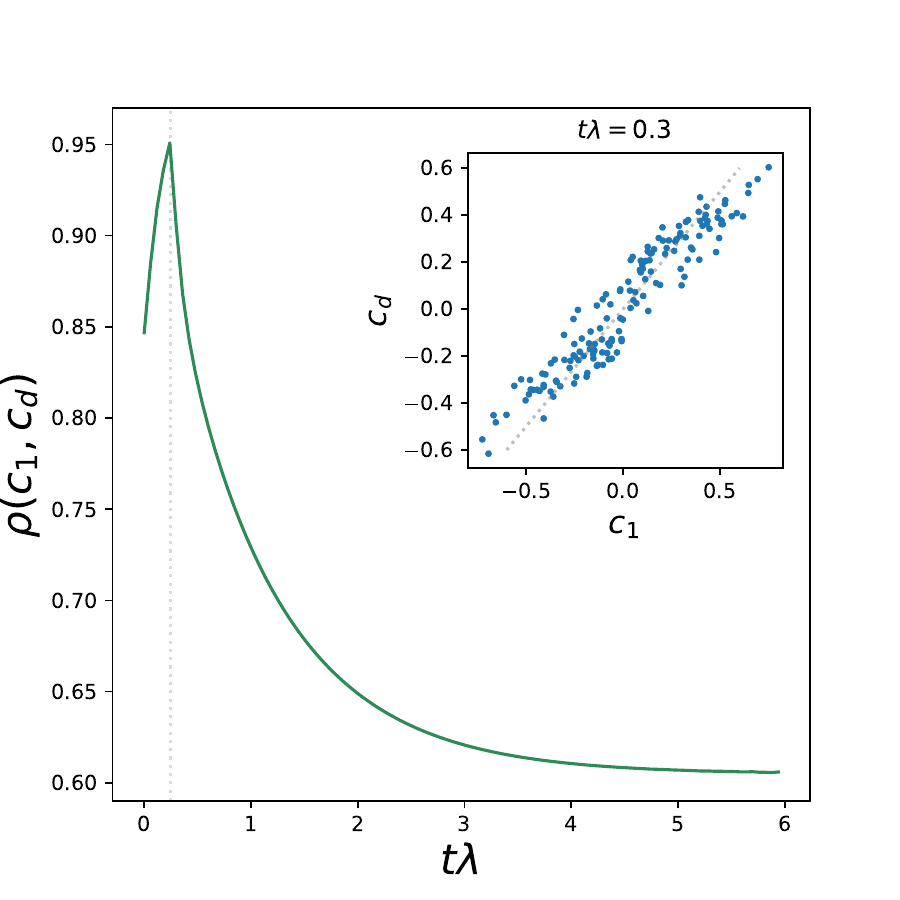}%
        \label{}%
    }
    \caption{Three relevant observables to benchmark the inferential performance of the unlearning regularization in time with $a = 0$: the standard deviation of the couplings $\sigma_J$, the Kullback-Leibler divergence between the original SK model and the inferred one at $\beta = 1$, the Pearson coefficient between the data correlation matrix $c_d$ and $c_1$, i.e. the correlation matrix for the model at $\beta = 1$. Choice of the parameters: $N = 18$, $\lambda = 0.06$, $\beta_d = 0.4$.}
    \label{fig:UNLa0_1}
\end{figure*}

We will now study the specific limit of the unlearning regularization that leads to an algorithm which strongly resembles the HU routine. The inferential performance of the learning procedure is examined by measuring the distance between the inferred model and the original one at different training steps. Results show, for the first time, that HU can be employed as an inferential tool, due to the beneficial effect of the Hebbian initialization of the couplings. We advance an explanation for its performance that is supported by further numerical evidence.
In the limit $a \rightarrow 0$ the updating rules for the parameters transform into Eqs. ~(\ref{eq:Juprob1}) and (\ref{eq:huprob1}).
As mentioned before, Eq.~(\ref{eq:Juprob1}) strongly resembles the traditional HU algorithm of Ref.~\cite{hopfield_unlearning_1983}. By contrast with the original rule, the new regularization samples configurations at $\beta = 1$ instead of stable fixed points of the neural dynamics, and makes use of a thermal average, rather than summing each contribution each time. 
We will keep dealing with small networks so that, given the initial conditions for the parameters, the loss function of the problem can be minimized exactly. 
As a numerical experiment, we train a BM with $N = 18$ neurons to learn a SK model at $\beta_d = 0.4$. The unlearning regularization is applied with $a = 0$ and the usual initial conditions for the parameters, i.e. $\boldsymbol{J}^{(0)} = \boldsymbol{c}_d$ and $\boldsymbol{h}^{(0)} = \boldsymbol{0}$.

Even if training is performed over both the couplings and the fields, according to the rules (\ref{eq:Juprob1}) and (\ref{eq:huprob1}), the fields $\boldsymbol{h}$ do not appear in the energy of the original model, hence we will focus exclusively on the evolution of the interactions $\boldsymbol{J}$.   
Fig. \ref{fig:UNLa0_1}a displays the standard deviation of the couplings: as we can see the general trend shows an exponential decay of the interactions, as one typically observes in the HU algorithm \cite{van_hemmen_increasing_1990}. The decay to zero is implied by the fact that the fixed point of the gradient descent equations for Eq.~\eqref{eq:Juprob1} corresponds to $\boldsymbol{c}_1 \equiv 0$. 
Fig. \ref{fig:UNLa0_1}b depicts the symmetric Kullback-Leibler divergence between the Gibbs-Boltzmann distribution of the original SK model, from which data were sampled, and the inferred model with $\beta = 1$. The $sD_{KL}$ is high at the beginning, then decreases until reaching a global minimum signaling an optimum in the inferential performance, then it increases again and stabilizes at a plateau. The minimum is sufficiently low to indicate a good statistical consistency between the model and data. 
The Pearson coefficient between the correlation matrix of the data $\boldsymbol{c}_d$ and the one of the inferred model at unitary temperature $\boldsymbol{c}_1$ is measured step-by-step in the learning process and reported in \cref{fig:UNLa0_1}c: as one can notice, there is a global maximum near the position of the global minimum of the $sD_{KL}$. 
The inset in \cref{fig:UNLa0_1}c displays the good agreement between the two matrices, the inferred $\boldsymbol{c}_1$ and $\boldsymbol{c}_d$, at the position of the peak.

This analysis suggests that the unlearning regularization with $a = 0$ displays two algorithmic regimes: a transient regime where the model at $\beta = 1$ shows a good statistical agreement with the generating model; another regime where $\boldsymbol{c}_1 \rightarrow 0$, and the total performance deteriorates. 
The first transient regime is the most important one, because it suggests that HU can be considered as an inferential tool, aside of its associative memory use. We know from \cref{sec:BML} that in BM learning data are shown to the model each time-step of the algorithm and this imposes the moment matching.
The co-existence of a positive Hebbian term and a negative unlearning one in the traditional BM learning has already been pointed out in Refs.~\cite{hinton_unsupervised_1999, hinton_forward-forward_2022}.
\begin{figure*}[t]
    \subfloat[]{%
        \includegraphics[width=.5\linewidth]{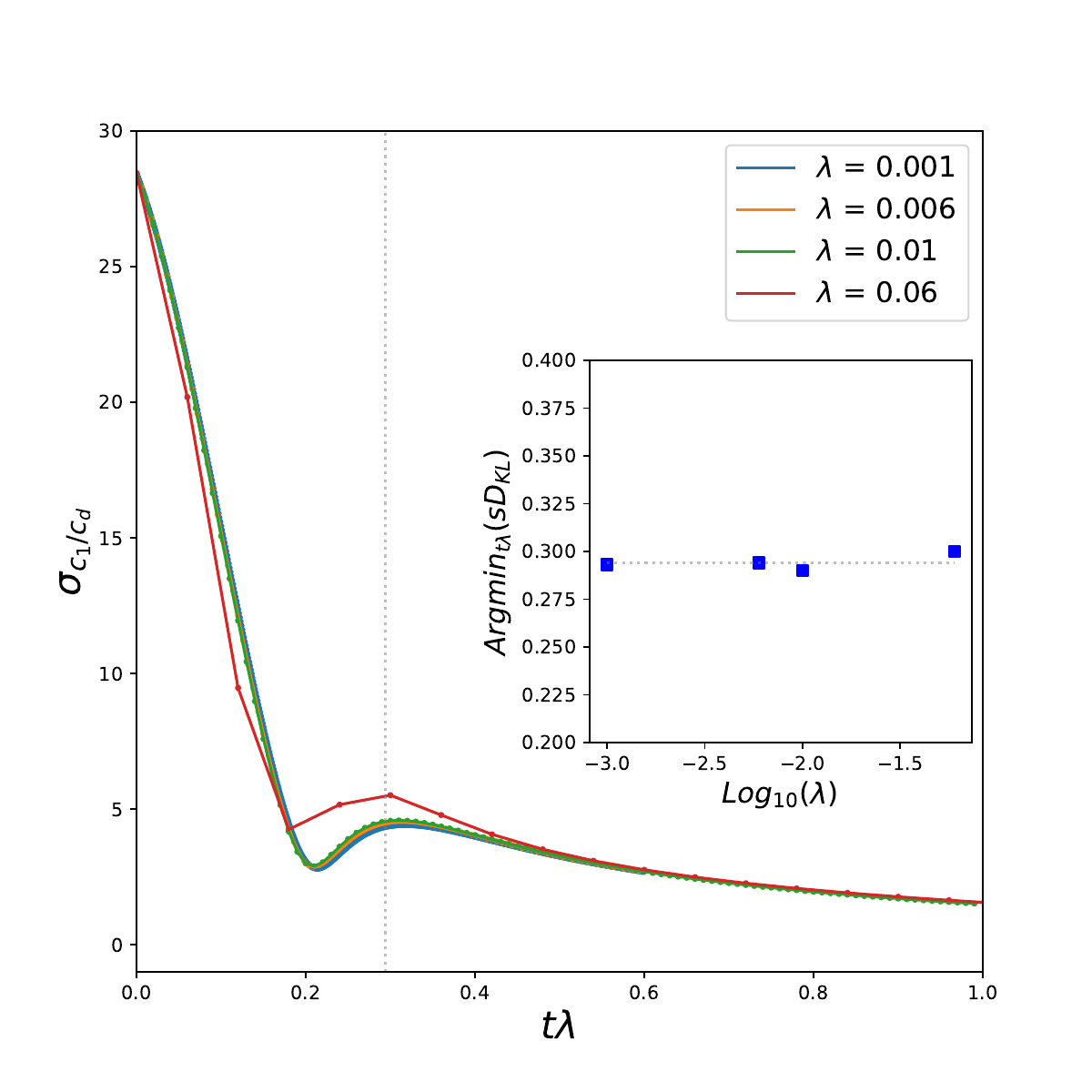}%
        \label{}%
    }\hfill
    \subfloat[]{%
        \includegraphics[width=.5\linewidth]{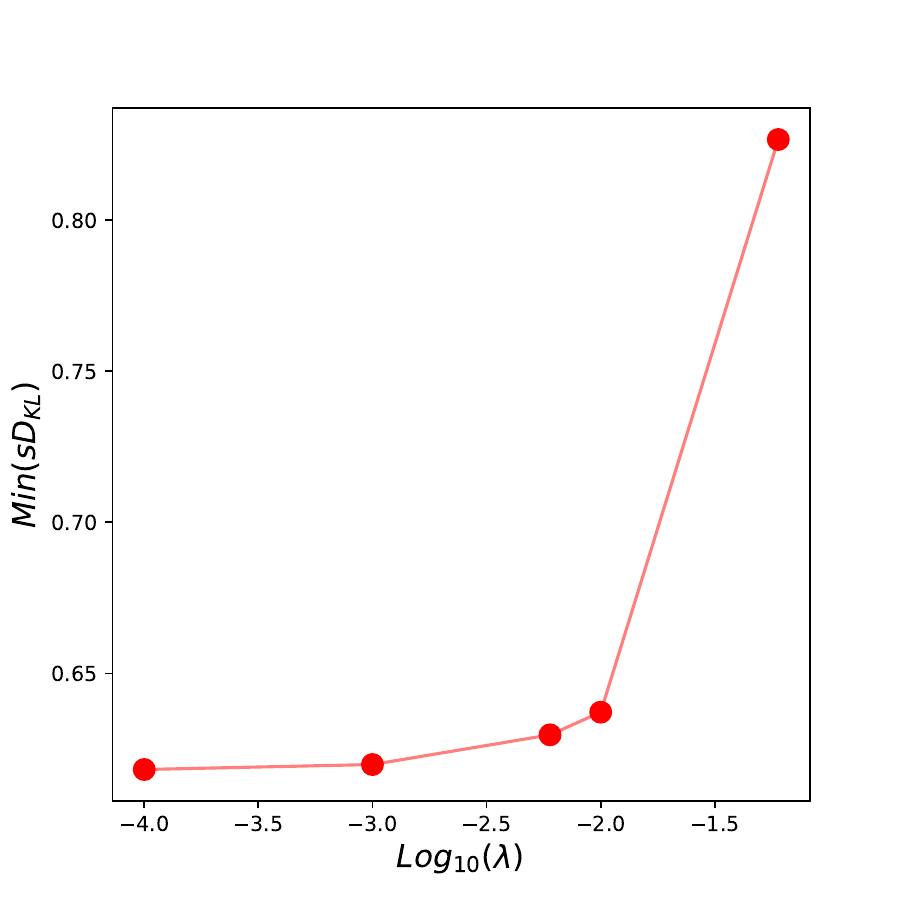}%
        \label{}%
    }
    \caption{(a) Standard deviation of the ratio between the elements of $\boldsymbol{c}_1$ and $\boldsymbol{c}_d$ as a function of the normalized time $t\lambda$; the subplot reports the values of $t\lambda$ associated to global minima of the symmetric Kullback-Leibler divergence ($sD_{KL}$) between the generating model and the inferred one for different choices of $\lambda$; both the vertical and horizontal gray dotted lines report the mean of the points in the inset. (b) Global minimum of the symmetric Kullback-Leibler divergence between the generating model and the inferred one as a function of $\lambda$ in log-scale. Choice of the parameters: $N = 18$, $a = 0$, $\beta_d = 0.4$.}
    \label{fig:UNLa0_ratio}
\end{figure*}
In fact, the variation of each pair of couplings $J_{ij}$ is given by
\begin{equation}
    \label{eq:gradJ_new}
    \delta J_{ij} = \delta^H J_{ij} + \delta^U J_{ij}, 
\end{equation}
where we neglected the dependence on time, and
\begin{equation}
    \label{eq:deltaJ_H}
    \delta^H J_{ij} = \langle S_i S_j \rangle_{data}, 
\end{equation}
\begin{equation}
    \label{eq:deltaJ_U}
    \delta^U J_{ij} = -\langle S_i S_j \rangle_{mod}.
\end{equation}
We can interpret each step in the minimization of $\mathcal{L}$ as the result of two contributions, one constant quantity deriving from the data, and another one deriving from the evolving model. 
While a standard BM can start wherever in the space of the parameters and ends up at the fixed point of \cref{eq:gradJ_new}, in the HU case the update is performed by using only the negative unlearning contribution. Nevertheless, the data have been seen by the model, specifically through the Hebbian choice of the initial conditions. This observation suggests that HU approaches the minimum of the loss function in two temporally separated steps: by first descending along the \textit{data} direction and then, progressively, along the \textit{model} one. As a consequence, we can assume that such a two-step minimization is similar to standard BM learning as long as
\begin{equation}
    \label{eq:cond_2stepsBML}
    |\langle S_i S_j \rangle_{mod}| \gg |\langle S_i S_j \rangle_{data}|
\end{equation}
for most of the pairs $i,j$. To test whether this condition holds while training a BM regularized with $a = 0$, and initialized with $\boldsymbol{J}^{(0)} = \boldsymbol{c}_d$, we measure the standard deviation of the ratio between the elements  of $\boldsymbol{c}_1$ and $\boldsymbol{c}_d$ as a function of time.

\begin{figure*}[ht!]
    \subfloat[]{%
        \includegraphics[width=.33\linewidth]{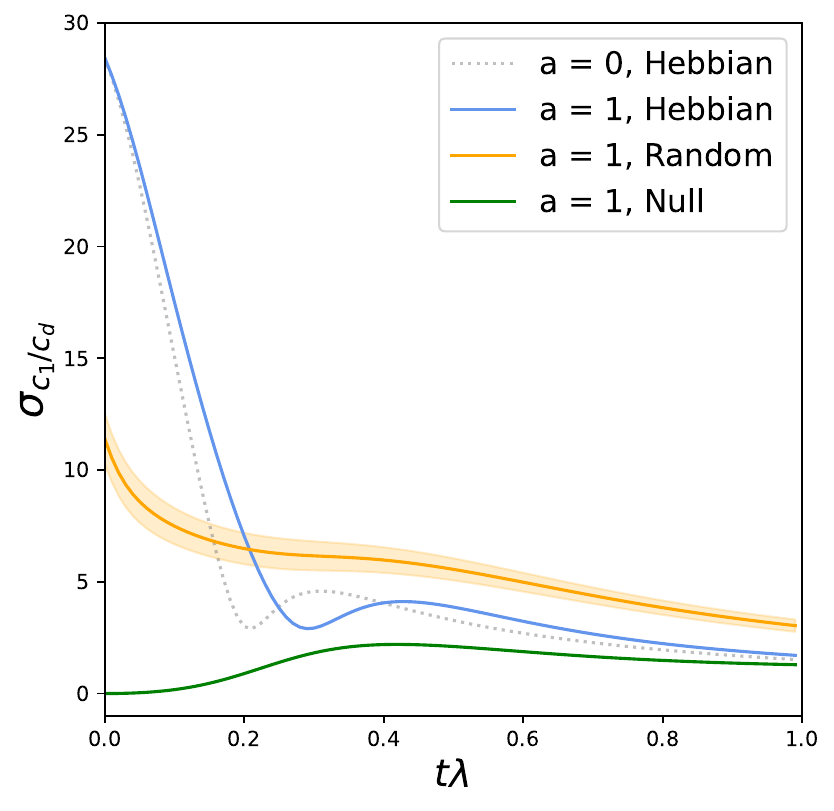}%
        \label{}%
    }\hfill
    \subfloat[]{%
        \includegraphics[width=.33\linewidth]{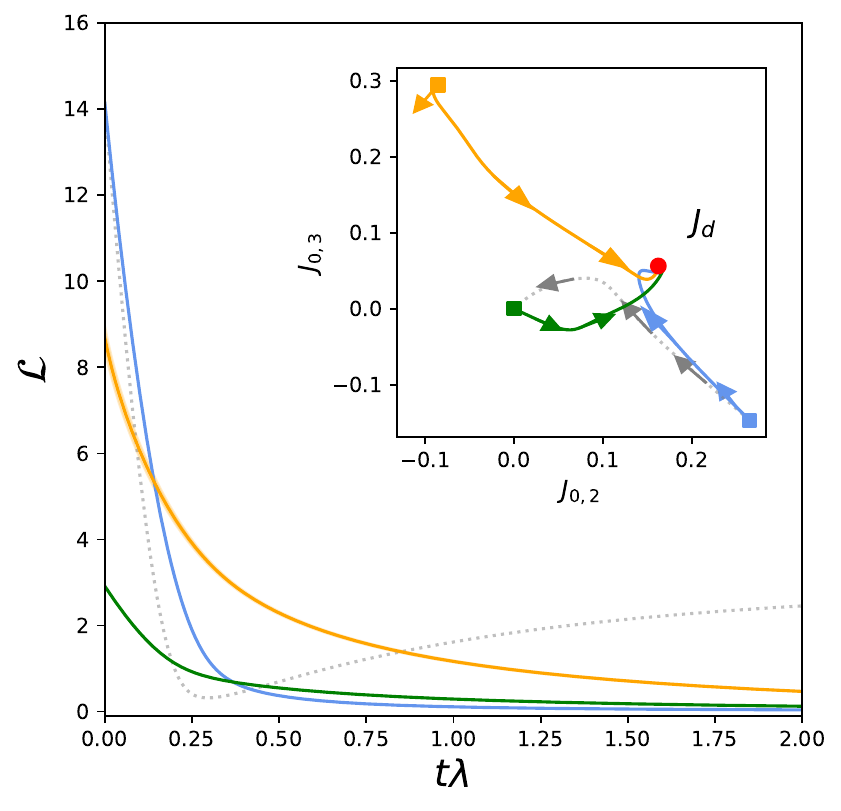}%
        \label{}%
    }\hfill
    \subfloat[]{%
        \includegraphics[width=.33\linewidth]{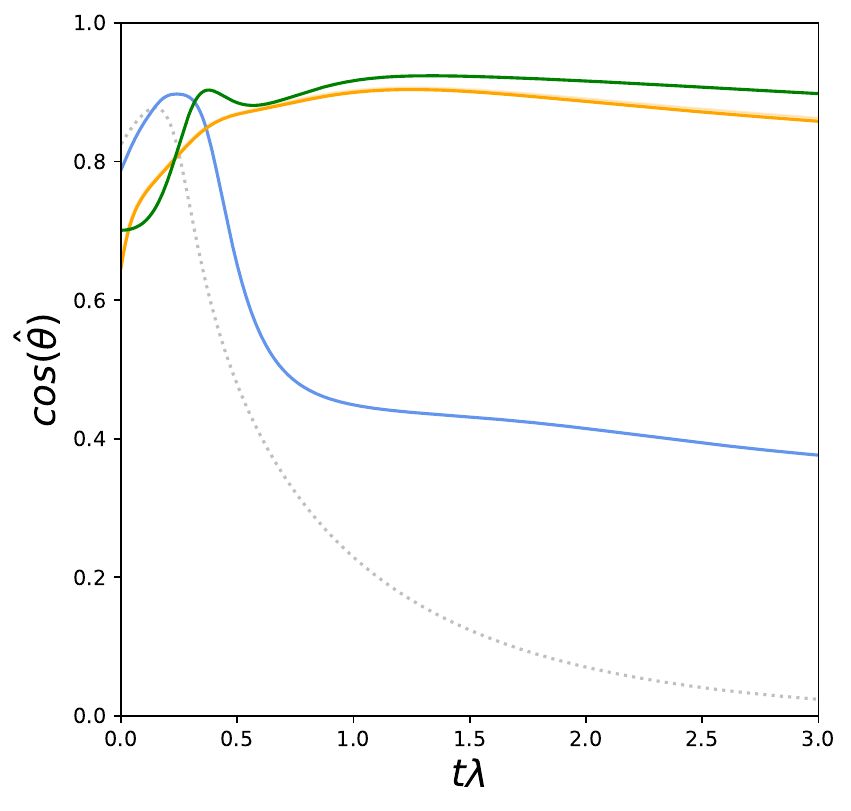}%
        \label{}%
    }
    \caption{Confronting different initializations of the parameters. (a) Standard deviation of the ratio between the elements of $\boldsymbol{c}_1$ and $\boldsymbol{c}_d$ as a function of the normalized time $t\lambda$ in a standard BM learning scenario (i.e. unlearning regularization with $a = 1$). (b) Loss function $\mathcal{L}$ as a function of the normalized time $t\lambda$ in a standard BM learning scenario. (c) Cosine of the angle between the gradient of $\boldsymbol{J}$ and the direction connecting the initial state of the parameters $\boldsymbol{J}^{(0)}$ and the model to be inferred $\boldsymbol{J}_d$. The curve for the unlearning regularization with $a = 0$ is indicated in all panels for comparison. The curve relative to the random initialization is averaged over fifteen starting matrices and the halo represents the standard deviation of the measures. Choice of the parameters: $N = 18$, $\beta_d = 0.4$, $\lambda = 0.01$.}
    \label{fig:UNLa0_ratio_initcomp}
\end{figure*}

Results are presented in \cref{fig:UNLa0_ratio}a for the usual data-set sampled by a SK model. The standard deviation of the ratio of the elements of the two matrices $\boldsymbol{c}_1$ and $\boldsymbol{c}_{d}$  is plotted as a function of $t \lambda$ for different choices of $\lambda$. The curves collapse well from $\lambda$ smaller than $\mathcal{O}(10^{-2})$. 
The system starts with $\boldsymbol{c}_1$ generally being much larger than $\boldsymbol{c}_d$, which satisfies the condition (\ref{eq:cond_2stepsBML}). Around $t \sim 0.2 \lambda^{-1}$ the curves reach a local minimum, then increase until a local maximum at $t \sim 0.29 \lambda^{-1}$, to start a slow decay right after. Both these two stationary points are located where $\boldsymbol{c}_1$ and $\boldsymbol{c}_d$ share the same order of magnitude, i.e. where the condition for the two-steps minimization of the loss ceases to hold. 
However, only the second stationary point is associated to the minimum of the $sD_{KL}$ between the original SK and the model, as reported in the inset of \cref{fig:UNLa0_ratio}a. Moreover, \cref{fig:UNLa0_ratio}b reports the values assumed by $sD_{KL}$ at its global minimum for various choices of $\lambda$. 
As one can notice, the generalization of the model increases when $\lambda$ is small: we can imagine that the smaller is $\lambda$ the stronger is the effect of the initial overshoot along the \textit{data} direction.

Different parameter initializations can now be compared. Specifically, we run the standard BM learning (i.e. the unlearning regularization with $a = 1$) after a Hebbian, random, and all-to-zero initialization of the couplings. The Hebbian and the random initializations have the same initial standard deviation of the couplings $\sigma_J$. 
At this stage we measure the evolution in time of the standard deviation of the ratio $\boldsymbol{c}_1/\boldsymbol{c}_d$ and the value of the loss. In addition to these two quantities, we introduce another observable that quantifies the degree of alignment of the gradient $\vec{\delta}J(t) = -\vec{\nabla}\mathcal{L}(t)$ to the direction connecting the initial state of the couplings, i.e. $\boldsymbol{J}^{(0)}$, and the convergence state $\boldsymbol{J}_d = \beta_d \boldsymbol{J}^{SK}$. Such quantity is defined as
\begin{equation}
    \label{eq:grad_projection}
    \cos{(\hat{\theta})} = \frac{\vec{\delta}J\cdot (\vec{J_d} - \vec{J}^{(0)})}{|\vec{\delta}J||\vec{J_d} - \vec{J}^{(0)}|}.
\end{equation}
The evolution of these observables are reported in \cref{fig:UNLa0_ratio_initcomp}. Note that, for the random initialization, measures are averaged over fifteen realizations of the initial matrix. As one can observe from \cref{fig:UNLa0_ratio_initcomp}a, the highest dispersion of the elements of $\boldsymbol{c}_1/\boldsymbol{c}_d$ is obtained through the Hebbian initialization; 
the random initialization also begins with the elements of $\boldsymbol{c}_1$ being on average larger than the empirical data correlations. Yet, as one can see from \cref{fig:UNLa0_ratio_initcomp}b, the loss decays generally faster in the Hebbian case, with respect to the random case. Figure \ref{fig:UNLa0_ratio_initcomp}c allows to interpret the behavior of the learning dynamics in terms of the gradient descent: 
the Hebbian initialization points the gradient $\vec{\delta}J$ towards the minimum of the loss, located in $\boldsymbol{J}_d$; on the other hand, the random initialization does not start well in its descent, and progressively adjusts the trajectory. The wrong start of training after random initialization can be also visualized in the inset of \cref{fig:UNLa0_ratio_initcomp}b, displaying a projection of the trajectory over the first two components of the $\boldsymbol{J}$ matrix, with arrows indicating the direction of the gradient descent. Only one experiment is reported for the random initialization since the others showed very similar behavior. 
Conversely, when the system learns from zero initial couplings, both the positive and negative contributions to the learning appear to be small and comparable in value, which implies that the implementation of pure unlearning is ineffective in this case; the descent of the loss is slower with respect to the Hebbian initialization; the trajectory is already sufficiently pointing to the bottom of the landscape, and progressively adjusts itself.

We conclude by stating another important remark. The unlearning trajectory and the standard BM learning look very similar in the first stage of training, in terms of the gradient descent, as predicted by Eq.~(\ref{eq:cond_2stepsBML}). After a certain amount of iterations that we have identified above, the unlearning trajectory deviates towards $\boldsymbol{J} = \boldsymbol{0}$, while the BM finally converges to the final target. The abrupt change of direction that is visible from both the two-dimensional projection of the trajectory and the evolution of the gradient might be related to the choice of the data-set.  
We have thus showed that, when the network is initialized in the Hebbian fashion, the dominant contribution to the BM learning is the unlearning one, and the gradient of $\boldsymbol{J}$ points towards the minimum of the loss function. Hence we showed that the HU algorithm works as well as an inferential device than as a training procedure for associative memory models.  

\section{\label{sec:neurons}A Real-World application}

This section is dedicated to the testing of the performance of the Unlearning regularization on real data. We train a BM on a data-set available from Ref.~\cite{schneidman_weak_2006}, reporting the activity of $N = 40$ retinal ganglion neurons reacting to visual stimuli.
We firstly binarize each data vector by discretizing a time interval of $0.1s$ into bins $\Delta \tau = 20ms$, as done in \cite{schneidman_weak_2006, mora_are_2011}. 
As a consequence, each data-point is a sequence of $S_i$ units, representing the instance where the $i-th$ neuron has fired $(S_i = +1)$ or not fired $(S_i = -1)$ in given time bin. 
Each time bin is thus associated to a data-point of our binary data-set. We are provided with several repetitions of the experiment over the same population of neurons, so that we can train the model with $M = 10^5$ data-points in total. 

\begin{figure}[t]
\centering
\includegraphics[width=0.8\linewidth]{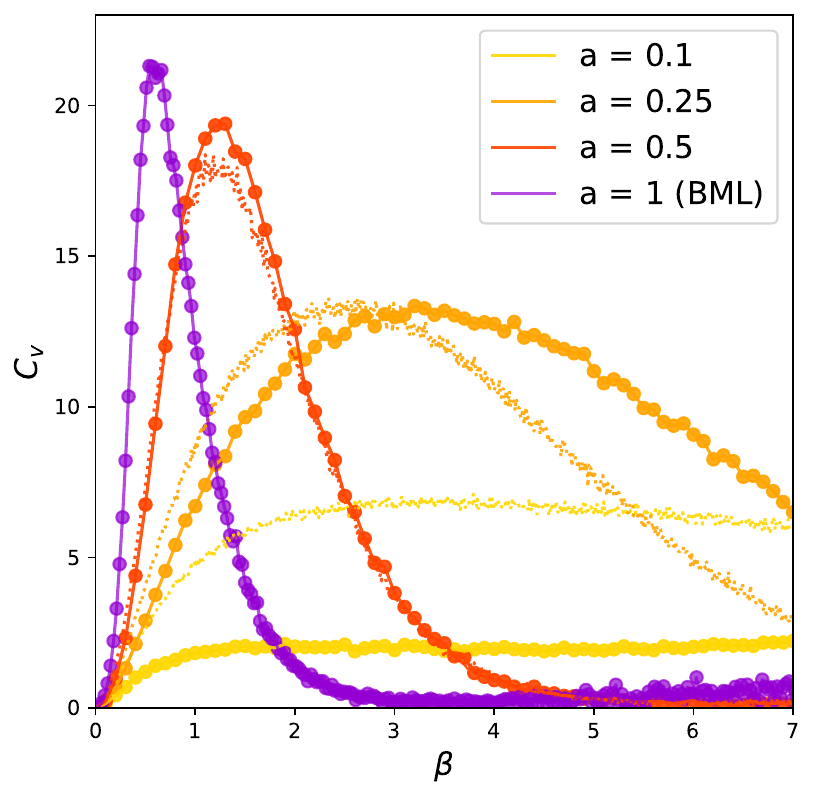}
\caption{Specific heat of a model inferred from the activity of $N = 40$ real neurons as a function of the inverse temperature $\beta$. The circles report the measure at convergence of the Unlearning regularization; the dots report measures when the algorithm is initialized in a Hebbian fashion and early-stopped when $\omega \simeq 1$. Error bars are neglected for clarity purposes. Choice of the parameters: $N = 40$, $M = 10^5$, $\lambda = 10^{-3}$.}
    \label{fig:bialek}
\end{figure}

The model is trained as a BM and regularized over couplings and fields through the Unlearning method with different choices of the parameter $a$. The profile of the specific heat $C_v$ is measured as a function of $\beta$ and reported in Fig. \ref{fig:bialek}. Lines with dots represent a machine trained until convergence of the algorithm. The curve obtained at $a = 1$, corresponding to pure BM learning, appears consistent with the one reported in \cite{mora_are_2011}. For each $a < 1$ the peak of the specific heat shifts away from unit, and its value in $\beta = 1$ is lowered by the regularization, suggesting a higher stability of the inferred network under a rescaling of couplings and fields. The small dots report the specific heat measured under the usual early-stopped prescription when couplings and fields are initialized as in equations (\ref{eq:J0}), (\ref{eq:h0}): even when $N$ becomes large there is an optimum amount of iterations where the Pearson coefficient between $\boldsymbol{c}_d$ and $\boldsymbol{c}_1$ is high and the slope $\omega$ reaches unity. As underlined by the figure, at this state of the network the specific heat is still significantly lower than the one obtained with no regularization, corroborating the importance of a Hebbian initialization for reaching the optimal level of accuracy and robustness of the model.   

\section{Conclusions}

In this study we have shown that:
\begin{itemize}
    \item In the context of Boltzmann Machines, and specifically for what concerns the robustness under rescaling of the parameters of the inferred network, the standard $L_1$ and $L_2$ regularization techniques are outperformed by the unlearning regularization. This result is valid for both cases where training is stopped earlier, to maximize the similarity between the inferred model and the ground-truth, and at convergence of the algorithms.  
    \item A particular case of the unlearning regularization (i.e. for a vanishing parameter $a$) reproduces a thermally averaged Hebbian Unlearning rule, that shows good inferential capabilities when initialized in the standard Hebbian fashion. Hebbian Unlearning can be interpreted as a two-steps Boltzmann-Machine learning and it is well reproduced by a $L_2$ regularization with a high regularization rate. In this case, our study stresses the necessity of early-stopping the training procedure after a specific amount of iterations where the performance reaches its optimum.
\end{itemize}

The goal of this article was thus to  show the effectiveness of unlearning as a form of regularization in Boltzmann Machines. 
The analysis has been carried out on both small networks, that allowed to compute all quantities without error by exact enumeration, and large networks, generally affected by statistical noise. For the most part of our work the ground-truth distribution was defined by one of two models, either the Curie-Weiss or the Sherrington-Kirkpatrick models, which are known to show a phase transition at a critical temperature. 
As a result, we could not only test the proximity of the inferred model to the original one, but also the susceptibility of the network under variations of the parameters, by comparing the trend of the specific heat of the ground truth model with respect to the inverse temperature, with the trend obtained after inference with regularization. 
In addition to this controlled framework, a real-world case, with data sampled from real neuronal activity, has been considered to confirm the results. Both heuristic and empirical arguments support the idea that models inferred via Boltzmann Machine learning tend to be pathologically critical \cite{schneidman_weak_2006, mora_are_2011, mastromatteo_criticality_2011}, i.e. less effective in sampling new examples. On the other hand, the regularized model is much less critical than the one originally learned in \cite{schneidman_weak_2006} from the same data-set.
Furthermore, we have displayed that even in the high-dimensional case one can characterize the performance of the algorithm in time by tracking two quantities: the Pearson coefficient and the slope of the empirical covariances versus the thermal correlations. 
These two observables are fast to compute and allow to establish an early stopping criterion that, combined with a Hebbian initialization of the parameters, reaches a greater generalization and robustness performance.
We conclude that, given a high consistency between the model and the ground-truth distribution, the stability of the model under rescaling of the parameters is enhanced by the unlearning regularization. 
As an alternative strategy to the unlearning regularization presented here, we also evaluated the minimization of the KL distance between the models at $\beta = 1$ and $\beta = a$. Even if the gradient descent equations look similar to the current case, they contain the contribution of four-point correlations among spins, resulting in a significant increase of the computational cost of the training. 


We also analyzed the particular limit of the regularization that gives as learning rules the ones expressed in equations (\ref{eq:Juprob1}) and (\ref{eq:huprob1}). 
The resulting procedure strongly resembles the traditional Hebbian Unlearning routine, but with a thermal two-point correlation computed at $\beta = 1$ replacing that computed on the zero-temperature fixed points of the dynamics. Previous work~\cite{nokura_paramagnetic_1996} supports the fact that the associative memory performance of such a \textit{thermal} unlearning does not deviate much from the original rule. Unlearning is not new to be mentioned in the statistical inference realm \cite{A1, A2, hinton_forward-forward_2022}, yet we are the first to consider the original rule proposed by Hopfield in \cite{hopfield_unlearning_1983}. 
As a new contribution to the characterization of unlearning, we tested its inferential power, i.e. the capability of learning the original model. 
So far, to the best of our knowledge, the original unlearning algorithm was exclusively tested in an associative memory framework, where data are actually patterns, that are sub-dominant in number with respect to the entire set of possibile configurations. 
Our results show a good inferential performance of the algorithm. The explanation for this success relies on the choice of the initialization of the parameters. A Hebbian initialization of the network has two implications. First, the contribution to the gradient of the loss deriving from the data (i.e. the positive \textit{Hebbian} contribution in the Boltzmann Machine learning in \cref{eq:Jup}) is much smaller than the contribution deriving from the cross-entropy of the model given the data (i.e. the negative \textit{unlearning} contribution in the Boltzmann Machine learning). This condition holds until an optimal amount of iterations that can be estimated from the numerical simulations.
Second, we showed that a Hebbian initialization of the couplings points the gradient of the parameters towards the correct direction, i.e. the one of the ground-truth of the inferential problem, at variance with other kinds of parameter initialization.
The beneficial effects of a Hebbian initialization for these type of machines has already emerged in the very recent literature \cite{pozas-kerstjens_efficient_2021,benedetti2}. Yet, the topological properties of the Hebbian free energy landscape that make them such a good starting point for training have not received due attention.

We conclude that Boltzmann Machine learning can be performed in two steps: an initial overshooting along the \textit{direction of the data} and then a gradual adjustment along the \textit{direction of the model} with increments being updated at each step of the algorithm. 
The effectiveness of a two-step training for a Boltzmann Machine encourages a biological interpretation of this kind of training, in agreement with the hypothesis that synapses in the brain could be fine tuned through the alternation of daily online experience and offline sleep \cite{sleep, girardeau_brain_2021}. Such a conjecture is inspiring members of the A.I. community in the development of training procedures for deep neural networks that might substitute the slow and non-biological back-propagation techniques \cite{hinton_forward-forward_2022, equilibrium_prop}. 
These observations, together with the importance of the Hebbian initialization of the network, appear to be consistent with previous results in literature concerning the unlearning algorithm and its implementation in associative memory modeling \cite{benedetti_supervised_2022, van_hemmen_increasing_1990, benedetti2,pozas-kerstjens_efficient_2021}.

\begin{acknowledgments}
We thank Marco Benedetti, Matteo Bisardi, Simone Ciarella, Sabrina Cotogno, Gabrielle Girardeau, Enzo Marinari, Felix Cosmin Mocanu, Giancarlo Ruocco, Beatriz Seoane, Martin Weigt and Jeanne Trinquier for many useful discussions. 
\end{acknowledgments}


\phantomsection 
\nocite{*}
\bibliographystyle{unsrt}
\bibliography{biblio}

\end{document}